\documentclass{jfm}

\usepackage{graphicx}
\usepackage{natbib}
\usepackage{caption}
\usepackage{subcaption}
\usepackage{hyperref}
\usepackage{color}
\usepackage{soul}

\ifCUPmtlplainloaded \else
  \checkfont{eurm10}
  \iffontfound
    \IfFileExists{upmath.sty}
      {\typeout{^^JFound AMS Euler Roman fonts on the system,
                   using the 'upmath' package.^^J}%
       \usepackage{upmath}}
      {\typeout{^^JFound AMS Euler Roman fonts on the system, but you
                   dont seem to have the}%
       \typeout{'upmath' package installed. JFM.cls can take advantage
                 of these fonts,^^Jif you use 'upmath' package.^^J}%
      }
  \else
  \fi
\fi

\ifCUPmtlplainloaded \else
  \checkfont{msam10}
  \iffontfound
    \IfFileExists{amssymb.sty}
      {\typeout{^^JFound AMS Symbol fonts on the system, using the
                'amssymb' package.^^J}%
       \usepackage{amssymb}%
       \let\le=\leqslant  
       \let\ge=\geqslant  
      }{}
  \fi
\fi

\ifCUPmtlplainloaded \else
  \IfFileExists{amsbsy.sty}
    {\typeout{^^JFound the 'amsbsy' package on the system, using it.^^J}%
     \usepackage{amsbsy}}
    {\providecommand\boldsymbol[1]{\mbox{\boldmath $##1$}}}
\fi

\usepackage[normalem]{ulem}

\newcommand{\fig}[1]{\textbf{Fig.~\ref{#1}}}
\newcommand{\eq}[1]{\textbf{Eq.~\ref{#1}}}
\newcommand{\sect}[1]{\textbf{\S\ref{#1}}}
\newcommand{\tbl}[1]{\textbf{Table~\ref{#1}}}

\renewcommand{\vec}[1]{\boldsymbol{#1}}
\newcommand{\bnabla}{\vec{\nabla}}
\newcommand{\zpos}{h}

\newcommand{\tfrac}[2]{\small{\frac{#1}{#2}}}
\newcommand{\textmds}[1]{\textmd{\tiny{#1}}}

\usepackage{array}
\newcolumntype{L}[1]{>{\raggedright\let\newline\\\arraybackslash\hspace{0pt}}m{#1}}
\newcolumntype{C}[1]{>{\centering\let\newline\\\arraybackslash\hspace{0pt}}m{#1}}
\newcolumntype{R}[1]{>{\raggedleft\let\newline\\\arraybackslash\hspace{0pt}}m{#1}}

\providecommand\bnabla{\boldsymbol{\nabla}}

\newsavebox{\astrutbox}
\sbox{\astrutbox}{\rule[-5pt]{0pt}{20pt}}

\newcommand{\ci}{\textmd{i}} 
\newcommand{\eqref}[1]{(\ref{#1})}

\title[Hydrodynamics of Micro-swimmers in Films]{Hydrodynamics of Micro-swimmers in Films}

\author[A.J.T.M. Mathijssen, A. Doostmohammadi, J.M. Yeomans and T.N. Shendruk]%
{A.J.T.M. Mathijssen$^1$%
  \thanks{Email address for correspondence: mathijssen@physics.ox.ac.uk},\ns
A. Doostmohammadi, J.M. Yeomans$^1$\break
and T.N. Shendruk$^1$}

\affiliation{$^1$Rudolf Peierls Centre for Theoretical Physics, University of Oxford, 1 Keble Road, Oxford, OX1 3NP, UK}

\pubyear{20xx}
\volume{0}
\pagerange{000--001}
\date{?; revised ?; accepted ?. - To be entered by editorial office}
\usepackage[latin1]{inputenc}
\begin{document}

\maketitle

\begin{abstract}
\noindent
One of the principal mechanisms by which surfaces and interfaces affect microbial life is by perturbing the hydrodynamic flows generated by swimming.
By summing a recursive series of image systems we derive a numerically tractable approximation to the three-dimensional flow fields of a Stokeslet (point force) within a viscous film between a parallel no-slip surface and no-shear interface and, from this Green's function, we compute the flows produced by a force- and torque-free micro-swimmer. 
We also extend the exact solution of Liron \& Mochon (1976) to the film geometry, which demonstrates that the image series gives a satisfactory approximation to the swimmer flow fields if the film is sufficiently thick compared to the swimmer size, and we derive the swimmer flows in the thin-film limit. 
Concentrating on the thick film case, we find that the dipole moment induces a bias towards swimmer accumulation at the no-slip wall rather than the water-air interface, but that higher-order multipole moments can oppose this. 
Based on the analytic predictions we propose an experimental method to find the multipole coefficient that induces circular swimming trajectories, allowing one to analytically determine the swimmer's three-dimensional position under a microscope.
\end{abstract}

\begin{keywords}
Authors should not enter keywords on the manuscript, as these must be chosen by the author during the online submission process and will then be added during the typesetting process (see http://journals.cambridge.org/data/\linebreak[3]relatedlink/jfm-\linebreak[3]keywords.pdf for the full list)
\end{keywords}

\newpage
%
\section{Introduction}
\label{sec:intro}
Beyond simply containing microbes and their surrounding fluids, surfaces and interfaces alter the behaviours, dynamics and even biological traits exhibited by swimming cells~\citep{bukoreshtliev13}.
Surfaces and interfaces affect micro-swimmer trajectories through hydrodynamics-induced interactions and impact the flows generated by microbes as they move through confined environments. 
The majority of research has focused on the effect of solid boundaries on swimming dynamics. 
In particular, the accumulation of bacteria at solid walls has been well demonstrated, both theoretically and in the experiments~ \citep{pedley87,lauga06,berke08,or09,or10,li09,drescher09,li11,spagnolie2012hydrodynamics,molaei2014failed, ishimoto2016simulation}. 
Likewise, recent studies have demonstrated the interplay between flowing fluids and swimming cells in various geometries for Newtonian~\citep{chacon2013chaotic,zottl2012nonlinear,zottl2013periodic,costanzo2012transport,masoud2013rotation,Dunkel2014,figueroa2015living} and non-Newtonian fluids~\citep{karimi2013hydrodynamic, ardekani2012emergence, mathijssen2015upstream}.

\begin{figure}
	\begin{center}
    	\includegraphics[trim = 0 30 0 0, clip,width=0.7\linewidth]{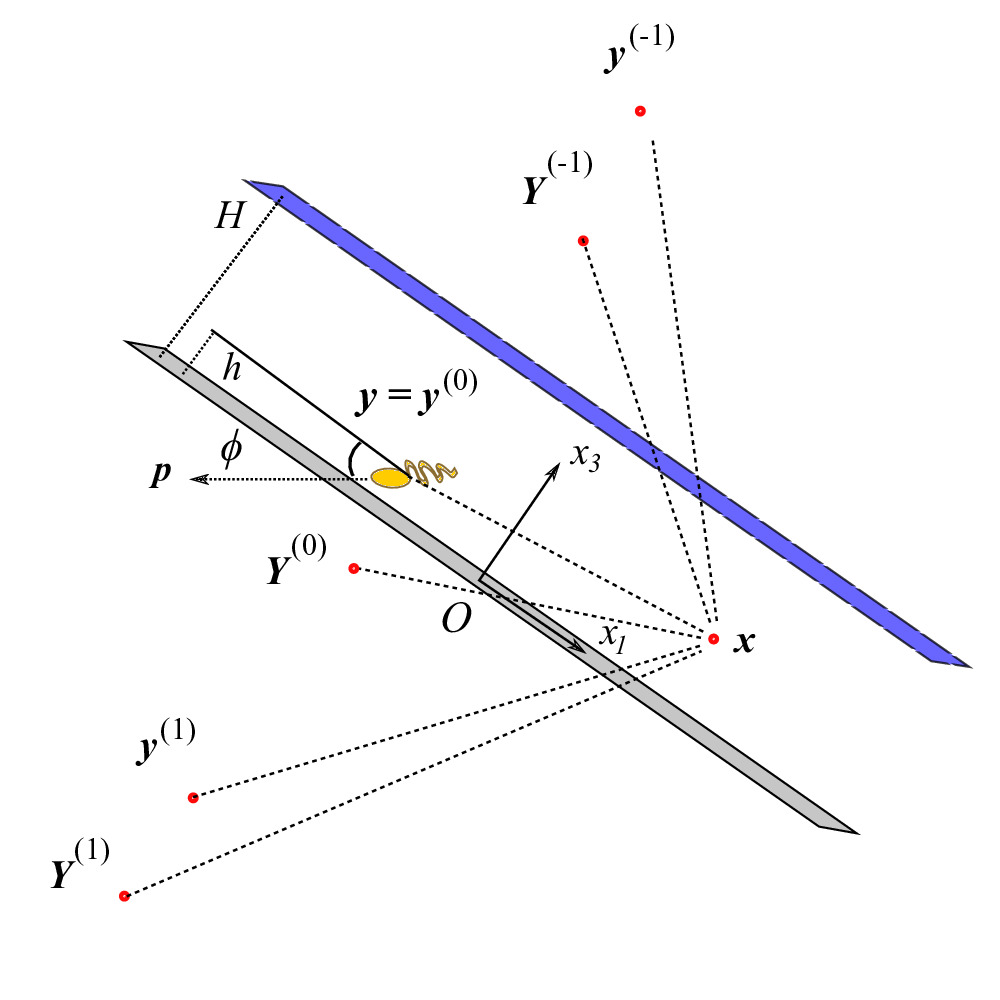}
	\end{center}
  	\caption{
Schematic showing the geometry of a micro-swimmer in a film at $\vec{y} = \vec{y}^{(0)}$ with orientation $\vec{p}$. 
Within the film micro-swimmers generate Stokesian flow fields that are found using a method of successive images at positions $\vec{y}^{(m)}, \vec{Y}^{(m)}$, shown by the red points. 
The swimmer and each of its infinite series of images contribute to the flow at any point $\vec{x}$. 
  	}
  	\label{fig:geometryDiagram}
\end{figure}

Less intently studied is the motion of micro-swimmers near fluid-fluid interfaces~\citep{Guasto2010,leonardo11,wang13,lopez14,masoud2014reciprocal,stone2015mobility} and, to the best of our knowledge, theoretical studies of motility in liquid films in contact with solid substrates (\fig{fig:geometryDiagram}) are rarely reported in the literature, despite their natural prevalence~\citep{Brandt2013, mathijssen2015hotspots}.
Innumerable habitats of small organisms are characterised as films that are macroscopically thin but substantially thicker than the characteristic size of swimming microbes and many experimental setups used for studies of various aspects of swimming cell dynamics essentially confine a culture of microorganisms between a substrate and liquid-liquid or liquid-gas interface. 
Liquid films allow for motility and swarming in order to colonise a wide variety of surfaces including plant and animal tissues~\citep{grimont1978genus, harshey1994dimorphic, bees2000interaction, harshey03}.
The motile microbial inoculant \textit{P. putida} traverses films as it moves through thin aquatic layers in soil~\citep{dechesne2010hydration}, as does \textit{E. coli}, which is known to swim upstream along crevices~\citep{hill07}.
\textit{P. syringae} bacteria can swarm on leaves by moving through their own secreted lubricant~\citep{quinones2005quorum}.
In fact, many bacteria secrete extracellular polymeric substances forming self-generated protective surface-bound biofilms~\citep{hall-stoodley04,givskov1997control,conrad12}, within which they then move. 

The flow fields generated by the propulsion of micro-swimmers have gathered a large interest from the fluid mechanics community \citep{blake1971note, liron1976Stokes, staben2003motion, crowdy2011two}
because they play an indispensable role in their ecological traits such as mechanosensing~\citep{AminPNAS, bukoreshtliev13}, energy expenditure~\citep{Guasto2010}, rheology~\citep{ishikawa2007rheology,guzman2012stochastic,gachelin2013non,lopez2015turning}, fluid mixing~\citep{kim2007controlled, leptos2009dynamics, ishikawa2010fluid, kurtuldu2011enhancement, mino2011enhanced, karimi2013gyrotactic, pushkin2014stirring, degraaf2016stirring, jeanneret2016entrainment} and nutrient uptake~\citep{magar2003nutrient, katija2012biogenic, jepson2013enhanced}.
Despite the widespread implications of swimming in films, the underlying hydrodynamics and its impact on the ecology of swimming cells has remained largely unexplored. 

In this paper, a detailed hydrodynamic description of swimmer dynamics within viscous films is developed by deriving the three-dimensional flows of a Stokeslet in a liquid film. We consider both a recursive series of image systems (\sect{sec:filmStokeslet}) and the exact solution using the method developed by \citet{liron1976Stokes}  (Appendix \sect{appsec:StokesletFlowFilm}).
A multipole expansion of the Stokeslet flow then gives the universal components of the flow field generated by a swimming micro-organism (\sect{sec:multipoleExpansion} and Appendix \sect{appsec:FarField}). 
Comparing the recursive series and the exact solution demonstrates their respective advantages and disadvantages in various regimes of film thickness  (\sect{sec:Comparison} and Appendix \sect{appsec:Comparison}). 
We conclude that the series solution is more amenable to a hydrodynamic multipole expansion and for numerically computing hydrodynamic interactions with the surfaces when the micro-swimmer is small compared to the film thickness.
In \sect{sec:swimmerDynamics} these results are used to predict the trajectories of ideal micro-swimmers. 
We explicitly map the dynamics and boundary accumulation of ideal cells defined by each successive hydrodynamic multipole moment (\sect{subsec:dipoleHI}--\sect{subsec:quadrupoleHI}), where the multipole parameters are directly linked to properties of the micro-organism, including size, shape and propulsion mechanism.
Together, these moments allow one to model more physical micro-organisms.
Employing our findings (\sect{subsec:rotletDoubletHI}), we propose an experimental method to determine a swimmer's rotary multipole coefficient, and its three-dimensional position under a microscope by measuring the radius of curvature of its projected trajectory.

\section{Flow fields in a film}
\label{sec:flows}

To derive the flow fields generated by a microbe swimming within a film constrained between a rigid wall and a free surface, we use a multipole expansion of the Stokes flow solution in a film. 
Performing such a multipole expansion requires a tractable analytic form of the hydrodynamic fields due to a force singularity (Stokeslet) within a viscous incompressible film of height $H$ and dynamic viscosity $\mu$. 
While the fundamental flow fields between two parallel boundaries have been found using a Fourier transform method \citep{liron1976Stokes}, we will present an analytical form that is particularly amenable to including the higher multipole moments required to accurately model micro-swimmers. 
This method is based on successive image reflections for finding the Stokeslet flow in a film. 
Previous studies have used a similar framework for studying the flow produced by mobile colloids \citep{ozarkar08}. 
To test our recursive series solution and establish the regimes where it is applicable, we derive the exact solution of the Stokeslet flow in a liquid film in Appendix \sect{appsec:StokesletFlowFilm}  by extending the method by \citet{liron1976Stokes}.

\subsection{Liquid film Stokeslet flow}
\label{sec:filmStokeslet}

The fluid is bounded by two parallel planar surfaces, a solid wall and an interface, at which no-slip and no-shear must respectively be satisfied, in addition to a no-penetration condition (\fig{fig:geometryDiagram}). 
We aim to solve the Stokes equations 
\begin{eqnarray}
\label{eq:StokesEqnA}
\vec{\nabla}  P(\vec{x}, t) - \mu \nabla^2 \vec{u}(\vec{x}, t)  &=& \vec{f} ~ \delta(\vec{x} - \vec{y}),
\\
\label{eq:StokesEqnB}
\vec{\nabla} \cdot \vec{u}(\vec{x}, t) &=& 0. 
\end{eqnarray}
Here, the fluid velocity $\vec{u}(\vec{x},t)$ and pressure $P(\vec{x},t)$ fields at location $\vec{x} = (x_1, x_2, x_3)$ and time $t$ are due to a point force $\vec{f}  ~ \delta(\vec{x} - \vec{y})$ (Stokeslet) that acts at position $\vec{y} = (y_1, y_2, y_3=h)$.
The boundary conditions for a film are the no-slip condition $\vec{u}(\vec{x}, t) = 0$ at the solid wall $x_3 = 0$ and both impermeability $u_3(\vec{x}, t) = 0$ and no-shear $\partial_3 \vec{u}(\vec{x}, t) = 0$ at the interface $x_3 = H$. 
In an unbounded fluid, the Green's function is the Oseen tensor \citep{kimmicrohydrodynamics}
\begin{eqnarray}
\label{eq:OseenTensor}
\mathcal{J}_{ij}(\vec{x},\vec{y}) &=& \frac{1}{8\pi \mu} \left( \frac{\delta_{ij}}{r} + \frac{r_i r_j}{r^3} \right), \qquad i,j \in \{1,2,3\}, 
\end{eqnarray}
where $\vec{r} = \vec{x} - \vec{y}$, $r=|\vec{r}|$ and $\delta_{ij}$ is the Kronecker delta.
The flow field is then given by $u_i^\textmds{S}(\vec{x},\vec{y},\vec{f}) = \mathcal{J}_{ij} f_j$, where repeated indices are summed over.
The corresponding pressure is $P(\vec{x},\vec{y},\vec{f}) = \mathcal{P}_j f_j$ with $ \mathcal{P}_j = r_j / 4\pi r^3$.

If the film height is taken to infinity ($H\rightarrow\infty$), then only the single no-slip boundary remains at $x_3=0$ and the Stokeslet flow field $\vec{u}^S$ is altered by the addition of an auxiliary flow field, that can be written in terms of a system of images. 
This image flow field is given by the Blake tensor \citep{blake1971note},
which is centered at the position $\vec{Y}^{(0)} = (y_1, y_2, -y_3) = \mbox{M} \cdot \vec{y}$, where the reflection matrix is $\mbox{M} = \mbox{diag}(1,1,-1)$.
The Blake tensor can be written in terms of the Oseen tensor~\citep[see][]{mathijssen2015tracer} as
\begin{eqnarray}
\label{eq:blakeTensor1}
\mathcal{B}_{ij} (\vec{x},\vec{Y}^{(0)})
&= 
(- \delta_{jk} + 2 y_3 \delta_{k3} \tilde{\partial}_j + y_3^2 \mbox{M}_{jk} \tilde{\nabla}^2) \mathcal{J}_{ik}(\vec{x}, \vec{Y}^{(0)}),
\end{eqnarray}
which is a function of $\vec{x}$ and $\vec{y}$, where the derivatives $\tilde{\partial}_j = \tfrac{\partial}{\partial y_j} = \mbox{M}_{jl} \tfrac{\partial}{\partial Y_l^{(0)}}$  
and $\tilde{\nabla}^{2} = \tilde{\partial}_l \tilde{\partial}_l$ 
are with respect to the force position $\vec{y}$. 
The tensor $\mathcal{B}_{ij} (\vec{x},\vec{Y}^{(0)})$ is recorded in the first row of \tbl{tab:images}. 
The resulting flow field at $\vec{x}$ due to a point force at $\vec{y} = \vec{y}^{(0)}$ in the vicinity of a no-slip wall is then 
$u^\textmds{B}_i = \left[ \mathcal{J}_{ij}(\vec{x},\vec{y}^{(0)}) + \mathcal{B}_{ij} (\vec{x},\vec{Y}^{(0)}) \right] f_j$. 

Similarly, if only a shear-free interface is present at $x_3=H$, the auxiliary flow field of the Stokeslet flow is a direct reflection centered at the position $\vec{Y}^{(-1)} = (y_1, y_2, 2H - y_3)$, with the corresponding free-slip boundary tensor
\begin{eqnarray}
\label{eq:freeBoundaryTensor}
\mathcal{T}_{ij} (\vec{x},\vec{Y}^{(-1)})
&=&
\mbox{M}_{jk} \mathcal{J}_{ik}(\vec{x}, \vec{Y}^{(-1)}).
\end{eqnarray}
This result is recorded in row two of \tbl{tab:images}. 

When both boundaries are present, the image system at $\vec{Y}^{(0)}$ corrects the boundary conditions at the solid no-slip wall ($x_3=0$) but disturbs the boundary conditions at the film interface ($x_3=H$), and vice versa for the image at $\vec{Y}^{(-1)}$. 
This difficulty can be overcome by using an infinite series of images to find the flow that satisfies the film Stokes equations. 
That is, the image system at $\vec{Y}^{(0)}$ (or $\vec{Y}^{(-1)}$) can be reflected in the interface (or wall) to form a secondary image system at position $\vec{y}^{(-1)}$ (or $\vec{y}^{(1)}$), and so on.
Hence, the positions of the image systems are 
\begin{eqnarray}
\label{eq:imagePositions1}
\vec{y}^{(m)} &=& (y_1, y_2, y_3 - 2mH),
\\
\label{eq:imagePositions2}
\vec{Y}^{(m)} &=& (y_1, y_2, - y_3 - 2mH),
\end{eqnarray}
where $m = 0, \pm1, \pm2, \dots$.

\fig{fig:geometryDiagram} schematically shows the series of images. 
As the number of images goes to infinity, the boundary conditions at both surfaces are satisfied. 
\tbl{tab:images} lists the procedure to find the image system tensors $\mathcal{G}_{ij}$, and hence the velocity fields, of the first few image systems of a Stokeslet in a film.
\tbl{tab:imagesExplicit} gives the resulting expressions of these tensors explicitly.
The tensor of a given image system can be obtained by replacing all the Oseen tensors $\mathcal{J}_{ij}$ in the tensor of the previous image system by the appropriate Blake tensor $\mathcal{B}_{ij}$ or free-slip boundary tensor $\mathcal{T}_{ij}$ given by \textbf{Eqs.~\ref{eq:blakeTensor1}-\ref{eq:freeBoundaryTensor}}, respectively.
It is key that all resulting expressions are still in terms of Oseen tensors and their derivatives, which can again be replaced at the next reflection operation.
Hence, by successively repeating the reflection operations (denoted by B or T for a `bottom' or `top' surface, operating from right to left), the image system tensor and thus the velocity field of the image systems is found via the recursion relations
\begin{eqnarray}
\label{eq:imageReplacementRule1}
\mathcal{G}_{ij}(\vec{x}, \vec{Y}^{(m)}) &=& \mbox{B} \; \mathcal{G}_{ij}(\vec{x}, \vec{y}^{(-m)}), \\
\label{eq:imageReplacementRule2}
\mathcal{G}_{ij}(\vec{x}, \vec{Y}^{(-m)}) &=& \mbox{T} \; \mathcal{G}_{ij}(\vec{x}, \vec{y}^{(m-1)}), \\
\label{eq:imageReplacementRule3}
\mathcal{G}_{ij}(\vec{x}, \vec{y}^{(m)}) &=& \mbox{B} \; \mathcal{G}_{ij}(\vec{x}, \vec{Y}^{(-m)}), \\
\label{eq:imageReplacementRule4}
\mathcal{G}_{ij}(\vec{x}, \vec{y}^{(-m)}) &=& \mbox{T} \; \mathcal{G}_{ij}(\vec{x}, \vec{Y}^{(m-1)}),
\end{eqnarray}
where $m\ge1$, and the reflection operators B and T act linearly on all the Oseen tensors $\mathcal{J}_{ij}$ present in the image system tensor $\mathcal{G}_{ij}$, as defined in \tbl{tab:images}.
The foundations of the recursion relations are 
\begin{eqnarray}
\label{eq:imageReplacementRuleBasis}
\mathcal{G}_{ij}(\vec{x}, \vec{Y}^{(0)}) &=& 
\mbox{B} \; \mathcal{J}_{ij}(\vec{x}, \vec{y}^{(0)})
= \mathcal{B}_{ij} (\vec{x},\vec{Y}^{(0)}),
\\
\label{eq:imageReplacementRuleBasis2}
\mathcal{G}_{ij}(\vec{x}, \vec{Y}^{(-1)}) &=& 
\mbox{T} \; \mathcal{J}_{ij}(\vec{x}, \vec{y}^{(0)})
= \mathcal{T}_{ij} (\vec{x},\vec{Y}^{(-1)}).
\end{eqnarray}
From these rules we obtain the Green's function in a film from the infinite series
\begin{eqnarray}
\label{eq:fundamentalTensorInFilm}
\mathcal{F}_{ij}(\vec{x}, \vec{y}) = \sum_{m=-\infty}^{\infty} \left[ \mathcal{G}_{ij}(\vec{x}, \vec{y}^{(m)}) + \mathcal{G}_{ij}(\vec{x}, \vec{Y}^{(m)}) \right], \quad
\end{eqnarray}
giving the film Stokeslet flow $u_i^\textmds{F}(\vec{x}, \vec{y}, \vec{f}) = \mathcal{F}_{ij} f_j$. 

This successive reflection method can also be used to construct the flow fields in more general confinement geometries, such as the flow bounded by two no-slip plates or the flow between a no-slip and a fluid-fluid (partial-slip) interface, by using the appropriate reflection operations, instead of those in \textbf{Eqs. \ref{eq:blakeTensor1}-\ref{eq:freeBoundaryTensor}}.
Furthermore, \citet{staben2003motion} showed that the flow field generated by a Stokeslet between two no-slip plates can be written as two Blake images and a rapidly decaying integral term. We anticipate that the same could achieved for the Stokeslet flow in a liquid film.

\begin{table*}
\begin{footnotesize}
\begin{tabular}{|l||l|l|l|}
\hline
(n) 	& Position		& Replace 									& with 
\\ \hline
(0) 	& $\vec{y}^{(0)}$	&  ---  	& $\mathcal{J}_{ij}(\vec{x}, \vec{y}^{(0)})$     \\ 
(1) 	& $\vec{Y}^{(0)}$	&  $\mbox{B} \mathcal{J}_{ij}(\vec{x}, \vec{y}^{(0)})$  	& $\mathcal{B}_{ij} (\vec{x},\vec{Y}^{(0)}) = (- \delta_{jk} + 2 y_3 \delta_{k3} \tilde{\partial}_j + y_3^2 \mbox{M}_{jk} \tilde{\nabla}^2) \mathcal{J}_{ik}(\vec{x}, \vec{Y}^{(0)})$     \\ 
(2) 	& $\vec{Y}^{(-1)}$	&  $\mbox{T} \mathcal{J}_{ij}(\vec{x}, \vec{y}^{(0)})$  	& $ \mathcal{T}_{ij} (\vec{x},\vec{Y}^{(-1)}) = \mbox{M}_{jk} \mathcal{J}_{ik}(\vec{x}, \vec{Y}^{(-1)})$     \\  
(3) 	& $\vec{y}^{(-1)}$	&  $\mbox{T} \mathcal{J}_{ij}(\vec{x}, \vec{Y}^{(0)})$  	& $\mbox{M}_{jk} \mathcal{J}_{ik}(\vec{x}, \vec{y}^{(-1)})$    \\ 
(4) 	& $\vec{y}^{(1)}$	& $\mbox{B} \mathcal{J}_{ij}(\vec{x}, \vec{Y}^{(-1)})$  	& $(- \delta_{jk} + 2 (2H-y_3) \delta_{k3} \mbox{M}_{jl} \tilde{\partial}_l + (2H-y_3)^2 \mbox{M}_{jk} \tilde{\nabla}^2) \mathcal{J}_{ik}(\vec{x}, \vec{y}^{(1)})$     \\ 
(5) 	& $\vec{Y}^{(1)}$	& $\mbox{B} \mathcal{J}_{ij}(\vec{x}, \vec{y}^{(-1)})$  	& $(- \delta_{jk} + 2 (2H+y_3) \delta_{k3} \tilde{\partial}_j + (2H+y_3)^2 \mbox{M}_{jk} \tilde{\nabla}^2) \mathcal{J}_{ik}(\vec{x}, \vec{Y}^{(1)})$     \\ 
(6) 	& $\vec{Y}^{(-2)}$	& $\mbox{T} \mathcal{J}_{ij}(\vec{x}, \vec{y}^{(1)})$  	& $\mbox{M}_{jk} \mathcal{J}_{ik}(\vec{x}, \vec{Y}^{(-2)})$     \\  
(7) 	& $\vec{y}^{(-2)}$	& $\mbox{T} \mathcal{J}_{ij}(\vec{x}, \vec{Y}^{(1)})$  	& $\mbox{M}_{jk} \mathcal{J}_{ik}(\vec{x}, \vec{y}^{(-2)})$    \\ 
(8) 	& $\vec{y}^{(2)}$	& $\mbox{B} \mathcal{J}_{ij}(\vec{x}, \vec{Y}^{(-2)})$  	& $(- \delta_{jk} + 2 (4H-y_3) \delta_{k3} \mbox{M}_{jl} \tilde{\partial}_l + (4H-y_3)^2 \mbox{M}_{jk} \tilde{\nabla}^2) \mathcal{J}_{ik}(\vec{x}, \vec{y}^{(2)})$     \\ 
(9) 	& $\vec{Y}^{(2)}$	& $\mbox{B} \mathcal{J}_{ij}(\vec{x}, \vec{y}^{(-2)})$  	& $(- \delta_{jk} + 2 (4H+y_3) \delta_{k3} \tilde{\partial}_j + (4H+y_3)^2 \mbox{M}_{jk} \tilde{\nabla}^2) \mathcal{J}_{ik}(\vec{x}, \vec{Y}^{(2)})$     \\ 
(10) 	& $\vec{Y}^{(-3)}$	& $\mbox{T} \mathcal{J}_{ij}(\vec{x}, \vec{y}^{(2)})$  	& $\mbox{M}_{jk} \mathcal{J}_{ik}(\vec{x}, \vec{Y}^{(-3)})$     \\  
(11) 	& $\vec{y}^{(-3)}$	& $\mbox{T} \mathcal{J}_{ij}(\vec{x}, \vec{Y}^{(2)})$  	& $\mbox{M}_{jk} \mathcal{J}_{ik}(\vec{x}, \vec{y}^{(-3)})$    \\ 
(12) 	& $\vec{y}^{(3)}$	& $\mbox{B} \mathcal{J}_{ij}(\vec{x}, \vec{Y}^{(-3)})$  	& $(- \delta_{jk} + 2 (6H-y_3) \delta_{k3} \mbox{M}_{jl} \tilde{\partial}_l + (6H-y_3)^2 \mbox{M}_{jk} \tilde{\nabla}^2) \mathcal{J}_{ik}(\vec{x}, \vec{y}^{(3)})$     \\ 
(13) 	& $\vec{Y}^{(3)}$	& $\mbox{B} \mathcal{J}_{ij}(\vec{x}, \vec{y}^{(-3)})$  	& $(- \delta_{jk} + 2 (6H+y_3) \delta_{k3} \tilde{\partial}_j + (6H+y_3)^2 \mbox{M}_{jk} \tilde{\nabla}^2) \mathcal{J}_{ik}(\vec{x}, \vec{Y}^{(3)})$     \\ 
~ \vdots & ~ \vdots    & ~ \vdots       & ~ \vdots     \\ \hline
\end{tabular}
\end{footnotesize}
\caption{Recursion relations for the successive image systems of a Stokeslet in a liquid film. 
The first image system of the Oseen tensor (\eq{eq:OseenTensor}) from reflection in the bottom wall is the Blake tensor (\eq{eq:blakeTensor1}), and the second image from reflection in the top interface is the mirrored Oseen tensor (\eq{eq:freeBoundaryTensor}). 
Subsequent image systems are obtained from further reflection operations with $\mbox{B}$ denoting the ``bottom'' (no-slip wall) and $\mbox{T}$ the ``top'' (no-shear interface), that operate linearly on all the Oseen tensor terms $\mathcal{J}_{ij}$ of the image system tensor $\mathcal{G}_{ij}$.}
\label{tab:images}
\end{table*}
\begin{table*}
\begin{footnotesize}
\begin{tabular}{|l||l|}
\hline
(n) & Image system tensor $\mathcal{G}_{ij}(\vec{x}, \vec{y}^{(m)} \mbox{ or } \vec{Y}^{(m)}) =$\\ \hline
(0) 	 
& $\mathcal{J}_{ij}(\vec{x}, \vec{y}^{(0)})$  
\\ 
(1) 	
&$(- \delta_{jk} + 2 y_3 \delta_{k3} \tilde{\partial}_j + y_3^2 \mbox{M}_{jk} \tilde{\nabla}^2) \mathcal{J}_{ik}(\vec{x}, \vec{Y}^{(0)}) $
\\ 
(2) 	
& $\mbox{M}_{jk} \mathcal{J}_{ik}(\vec{x}, \vec{Y}^{(-1)})$     
\\  
(3) 	
& $(- \delta_{jk} + 2 y_3 \delta_{k3} \tilde{\partial}_j + y_3^2 \mbox{M}_{jk} \tilde{\nabla}^2) \mbox{M}_{kl} \mathcal{J}_{il}(\vec{x}, \vec{y}^{(-1)}) $
\\ 
(4) 	
& $\mbox{M}_{jk} (- \delta_{kl} + 2 (2H-y_3) \delta_{l3} \mbox{M}_{ko} \tilde{\partial}_o + (2H-y_3)^2 \mbox{M}_{kl} \tilde{\nabla}^2) \mathcal{J}_{il}(\vec{x}, \vec{y}^{(1)})$
\\ 
(5) 	
& $(- \delta_{jk} + 2 y_3 \delta_{k3} \tilde{\partial}_j + y_3^2 \mbox{M}_{jk} \tilde{\nabla}^2) \mbox{M}_{kl} (- \delta_{lo} + 2 (2H+y_3) \delta_{o3} \tilde{\partial}_l + (2H+y_3)^2 \mbox{M}_{lo} \tilde{\nabla}^2) \mathcal{J}_{io}(\vec{x}, \vec{Y}^{(1)})$
\\ 
(6) 	
& $\mbox{M}_{jk} (- \delta_{kl} + 2 (2H-y_3) \delta_{l3} \mbox{M}_{kp} \tilde{\partial}_p + (2H-y_3)^2 \mbox{M}_{kl} \tilde{\nabla}^2) \mbox{M}_{lo} \mathcal{J}_{io}(\vec{x}, \vec{Y}^{(-2)})$
\\ 
(7) 	
& $(- \delta_{jk} + 2 y_3 \delta_{k3} \tilde{\partial}_j + y_3^2 \mbox{M}_{jk} \tilde{\nabla}^2) \mbox{M}_{kl} (- \delta_{lo} + 2 (2H+y_3) \delta_{o3} \tilde{\partial}_l + (2H+y_3)^2 \mbox{M}_{lo} \tilde{\nabla}^2) \mbox{M}_{op} \mathcal{J}_{ip}(\vec{x}, \vec{y}^{(-2)})$
\\ 
~ \vdots & ~ \vdots      \\ \hline
\end{tabular}
\end{footnotesize}
\caption{
Explicit expressions of the image system tensors $\mathcal{G}_{ij}$ of the first few image systems of a Stokeslet in a liquid film.
The indices $i,j,k,l,o,p \in \{1,2,3\}$ and repeated indices are summed over. 
Added together, these tensors yield the Green's function in a film (\eq{eq:fundamentalTensorInFilm}).
}
\label{tab:imagesExplicit}
\end{table*}

\subsection{Multipole expansion}
\label{sec:multipoleExpansion}

\begin{figure}
\begin{center}
\includegraphics[width=\linewidth]{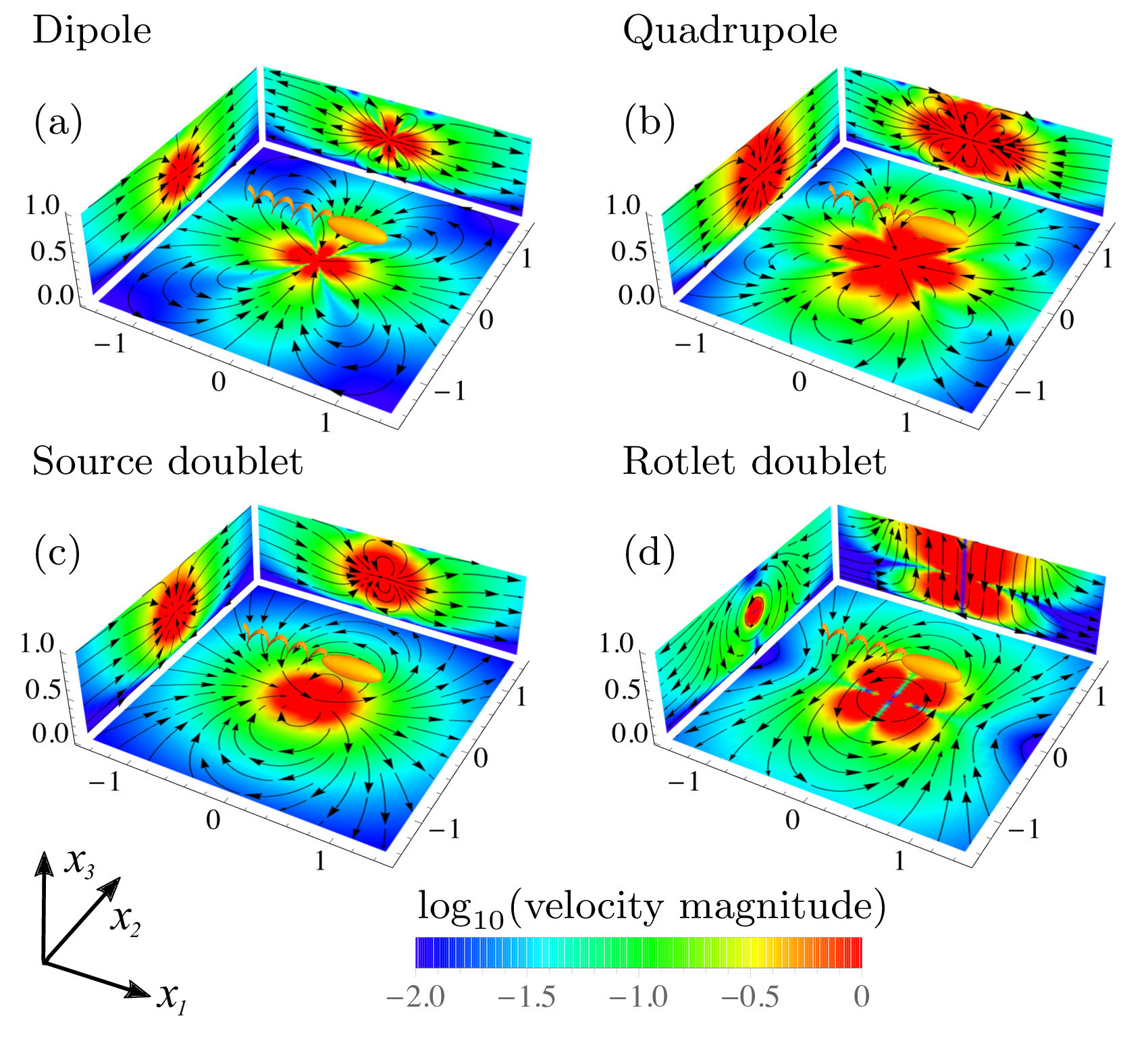} 
\end{center}
\caption{
Flow fields produced by a micro-swimmer at the centre of a film of height $H$, obtained using the recursive series method with $n=9$ images. 
Shown are
(a) Dipolar $\vec{u}^\textmds{D}$, \eq{eq:flowStokesDipole}, 
(b) Quadrupolar $\vec{u}^\textmds{Q}$, \eq{eq:flowStokesQuadrupole}, 
(c) Source doublet $\vec{u}^\textmds{SD}$, \eq{eq:flowSourceDoublet}, and 
(d) Rotlet doublet $\vec{u}^\textmds{RD}$, \eq{eq:flowRotletDipole}.
The flow fields shown correspond to planar cuts close to the swimmer, 
$x_1 = -a$, $x_2 = a$ and $x_3 = h-a$, where $h= H/2$, $a = H/100$ and all lengths in the figure are scaled with respect to $H$.
The schematic swimmer points in the swimming direction $\vec{p} = \vec{\hat{e}}_1$. 
Colormaps show the velocity magnitude normalised by its maximum on a logarithmic scale, ranging from $10^{-2}$ (blue) to $1$ (red), and are superimposed by streamlines (black lines). 
}
  \label{fig:freeSwimmerFlowFields}  
\end{figure}

In this section, we summarise how the flow field generated by a micro-swimmer is related to the Stokeslet in a film. 
The micro-organism is modelled as a prolate spheroid with semi-major and -minor axes $a$ and $b$, respectively and aspect ratio $\gamma=a/b$. 
Unless otherwise stated, we use $a$ as the characteristic swimmer size throughout the text.
The organism is located at position $\vec{y}$ with orientation $\vec{p}$, swimming in a film of height $H$ (\fig{fig:geometryDiagram}). In addition to its own motility (propulsion velocity $\vec{v}^\textmds{SW} = v^\textmds{SW} \vec{p}$), the motion of a swimming cell is affected by steric and hydrodynamic interactions with the bounding planes, plus any background flow. 

As a micro-swimmer moves it generates a flow $\vec{u}$. 
This swimmer-generated flow field can be written in terms of a multipole expansion of the Stokes flow solution in the film (\eq{eq:fundamentalTensorInFilm}).
Because neutrally buoyant micro-swimmers do not subject their surrounding fluid to a net force or torque, we exclude the Stokeslet and Stokes rotlet terms from the expansion. 
Similarly, the assumption that the swimmer is cylindrically symmetric about the swimming direction allows us to exclude the non-symmetric terms \citep[see][]{mathijssen2015tracer}.
Hence, an axisymmetric force-free and torque-free micro-swimmer generates a velocity field 
\begin{eqnarray}
 \label{eq:swimmerVelocityDef}
 \vec{u} \left(\vec{x}, \vec{y}, \vec{p} \right) &= \vec{u}^\textmds{D} + \vec{u}^\textmds{Q} + \vec{u}^\textmds{SD} + \vec{u}^\textmds{RD} + \ldots ,
\end{eqnarray}
\noindent where $\vec{u}^\textmds{D}$ is the Stokes dipole, $\vec{u}^\textmds{Q}$ the quadrupole, $\vec{u}^\textmds{SD}$ the source doublet, and $\vec{u}^\textmds{RD}$ the rotlet doublet \citep{spagnolie2012hydrodynamics, mathijssen2015tracer}.
Though including more than four terms in the multipole expansion might more accurately describe the near-field flow due to swimmer specific details, these few terms satisfactorily account for the universal attributes of a generic micro-swimmer. 

Each contribution to the multipole expansion can be written in terms of derivatives of the Green's function in the film (\eq{eq:fundamentalTensorInFilm}). Specifically,
\begin{eqnarray}
\label{eq:flowStokesDipole}
\vec{u}^\textmds{D}(\vec{x}, \vec{y}, \vec{p})  &=&  \kappa ~ (\vec{p} \cdot \tilde{\bnabla}) (8 \pi \mu \underline{\underline{\mathcal{F}}} \cdot \vec{p}),
\\
\label{eq:flowStokesQuadrupole}
\vec{u}^\textmds{Q}(\vec{x}, \vec{y}, \vec{p}) &=& - \tfrac{1}{2} \nu ~ (\vec{p} \cdot \tilde{\bnabla})^2 (8 \pi \mu \underline{\underline{\mathcal{F}}} \cdot \vec{p}),
\\
\label{eq:flowSourceDoublet}
\vec{u}^\textmds{SD}(\vec{x}, \vec{y}, \vec{p}) &=& - \tfrac{1}{2} \sigma ~ \tilde{\nabla}^2 (8 \pi \mu \underline{\underline{\mathcal{F}}} \cdot \vec{p}), 
\\
\label{eq:flowRotletDipole}
\vec{u}^\textmds{RD}(\vec{x}, \vec{y}, \vec{p}) &=& - \tfrac{1}{2} \tau ~ (\vec{p} \cdot \tilde{\bnabla}) \tilde{\bnabla} \times (8 \pi \mu \underline{\underline{\mathcal{F}}} \cdot \vec{p}),
\end{eqnarray}
where the derivatives act on the swimmer position $\vec{y}$. 
The multipole coefficients $\kappa$ and ($\nu$, $\sigma$, $\tau$) have dimensions of [velocity $\times$  length$^2$] and [velocity $\times$ length$^3$], respectively. 

These multipole flow fields (\textbf{Eqs. \ref{eq:flowStokesDipole}-\ref{eq:flowRotletDipole}}) are shown in \fig{fig:freeSwimmerFlowFields}. 
The dipole arises from the opposing propulsion and drag forces exerted by the swimmer ($\vec{u}^\textmds{D}$; \fig{fig:freeSwimmerFlowFields}a). 
Pusher-type swimmers such as \textit{E. coli} have a positive force dipole with $\kappa>0$ (shown in \fig{fig:freeSwimmerFlowFields}a). 
Pushers drive fluid out along the swimming direction and draw fluid in from the sides. 
Puller-type swimmers, on the other hand, have $\kappa<0$ and draw fluid in along their swimming axis~\citep{lauga2009hydrodynamics}.
The quadrupole flow field represents the fore-aft asymmetry of the microorganism ($\vec{u}^\textmds{Q}$; \fig{fig:freeSwimmerFlowFields}b). 
For example, the quadrupole describes the weighting of propulsion forces towards the posterior of flagellated bacteria, in which case one expects $\nu >0$. 
The source doublet represents the finite size of the swimmer ($\vec{u}^\textmds{SD}$; \fig{fig:freeSwimmerFlowFields}c).
For ciliated organisms with a slip velocity at their surface $\sigma>0$, whereas for non-ciliated swimmers one would expect $\sigma <0$ because this corresponds to the Fax\'{e}n correction to the Stokeslet flow for a finite-sized solid sphere.
Finally, the rotlet doublet represents the opposing rotation of the swimmer's head and tail ($\vec{u}^\textmds{RD}$; \fig{fig:freeSwimmerFlowFields}d) \citep{spagnolie2012hydrodynamics, mathijssen2015tracer}.


\begin{figure*}
\begin{center}
\includegraphics[width = \linewidth]{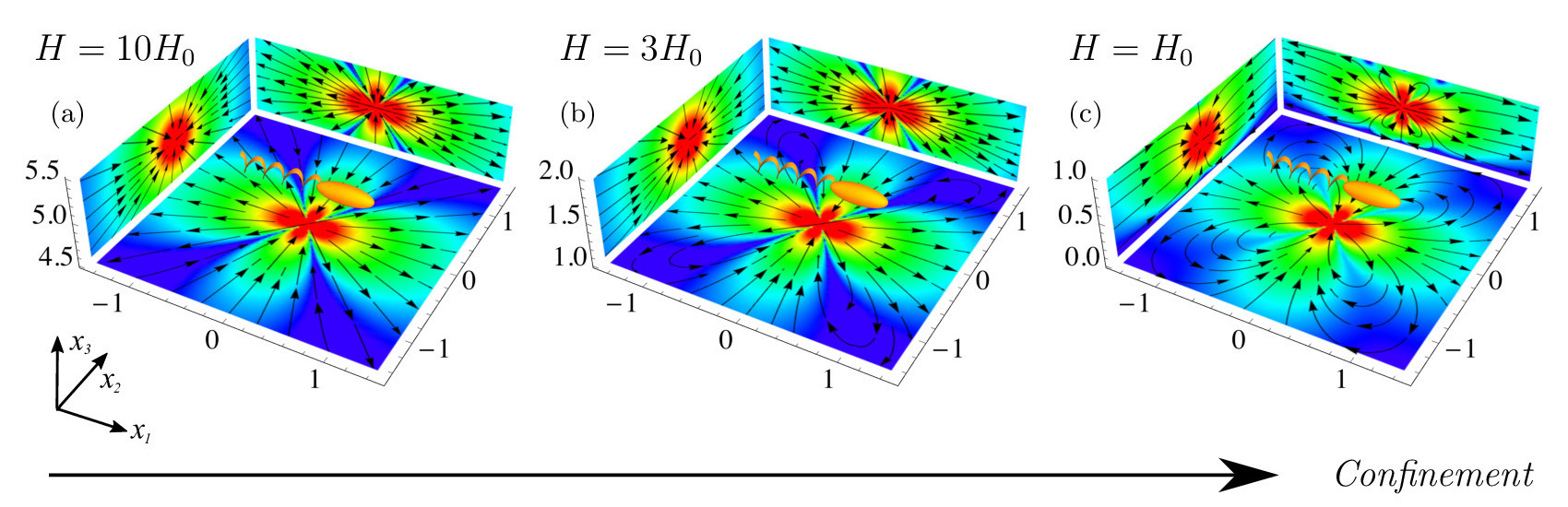}
\end{center}
\caption{The flow field generated by a micro-swimmer is modified in thin films. 
Panels (a,b,c) show flow fields for decreasing values of the film height, $H=10 H_0$, $H=3 H_0$, and $H=H_0$, respectively.
The micro-swimmer is located at the middle of the film, at $h=H/2$, and oriented in the direction $\vec{p} = \vec{\hat{e}}_1$.
Although the flow is purely dipolar far away from the boundaries (a), this lower-order multipole is screened with increasing confinement and a recirculating flow pattern appears (b,c).  
The number of images, planar cuts and colour legends are defined as in \fig{fig:freeSwimmerFlowFields}.
}
\label{fig:flow}
\end{figure*}

In a film of sufficiently large thickness the flow field is relatively unaffected by the boundaries (\fig{fig:flow}a).
Upon decreasing the height of the film, the flow profiles in the $x_1-x_3$ and $x_2-x_3$ planes remain unaffected, except near the surfaces where the boundary conditions must be satisfied. 
However, as the thickness of the film is reduced the flow field in the $x_1-x_2$ plane is modified and recirculating flow patterns appear close to the swimming cell (\fig{fig:flow}b). Such patterns are reminiscent of those seen for a Stokeslet between two parallel plates \citep{liron1976Stokes}.
The recirculating regions are enhanced in size as the film thickness is further reduced (\fig{fig:flow}c) compared to the size of the swimmer. 

The effect of reducing the film thickness on altering the flow structure is understood by considering the minimisation of the energy dissipation by a self-propelled organism in Stokes flow.
As the thickness of the film is reduced, the propulsion energy is more effectively dissipated by the boundaries. 
This introduces a length scale of order $\sim H$ beyond which the primary contributions to the flow field are screened. 
As a result of the emergence of this screening length, the flow field is suppressed in the plane normal to the film height and recirculating flow patterns are formed.

To understand the case of strong confinement, we derive the Stokeslet and swimmer-generated flow fields in the thin-film limit (Appendix \sect{appsec:FarField}).
In this limit, the flows in the $i$ direction due to a swimmer pointing in the $j$ direction decay exponentially with the lateral distance if either $i$ or $j$ or both are equal to three (\textit{i.e.} directed perpendicular to the film).
Only the parallel components of the flow do not decay exponentially, and those have a half-parabolic profile along $x_3$.
The Stokeslet has a recirculating pattern of two loops in the $x_1-x_2$ plane, the dipole has four loops, the quadrupole has six loops, and the source doublet maintains its bulk-flow structure with two loops (Appendix \sect{appsec:FarField}; \fig{appfig:ThinFilmFlows}).
The solution for higher order multipoles between two no-slip surfaces, which we have not found elsewhere in the literature, are the same, but with a parabolic profile along $x_3$. 
These thin-film limit expressions could be used, for example, to model swimmer-swimmer interactions in films of thickness comparable to the swimmer size.

\subsection{Comparison to the exact solution}
\label{sec:Comparison}

\begin{figure}
\begin{center}
\includegraphics[width=\linewidth]{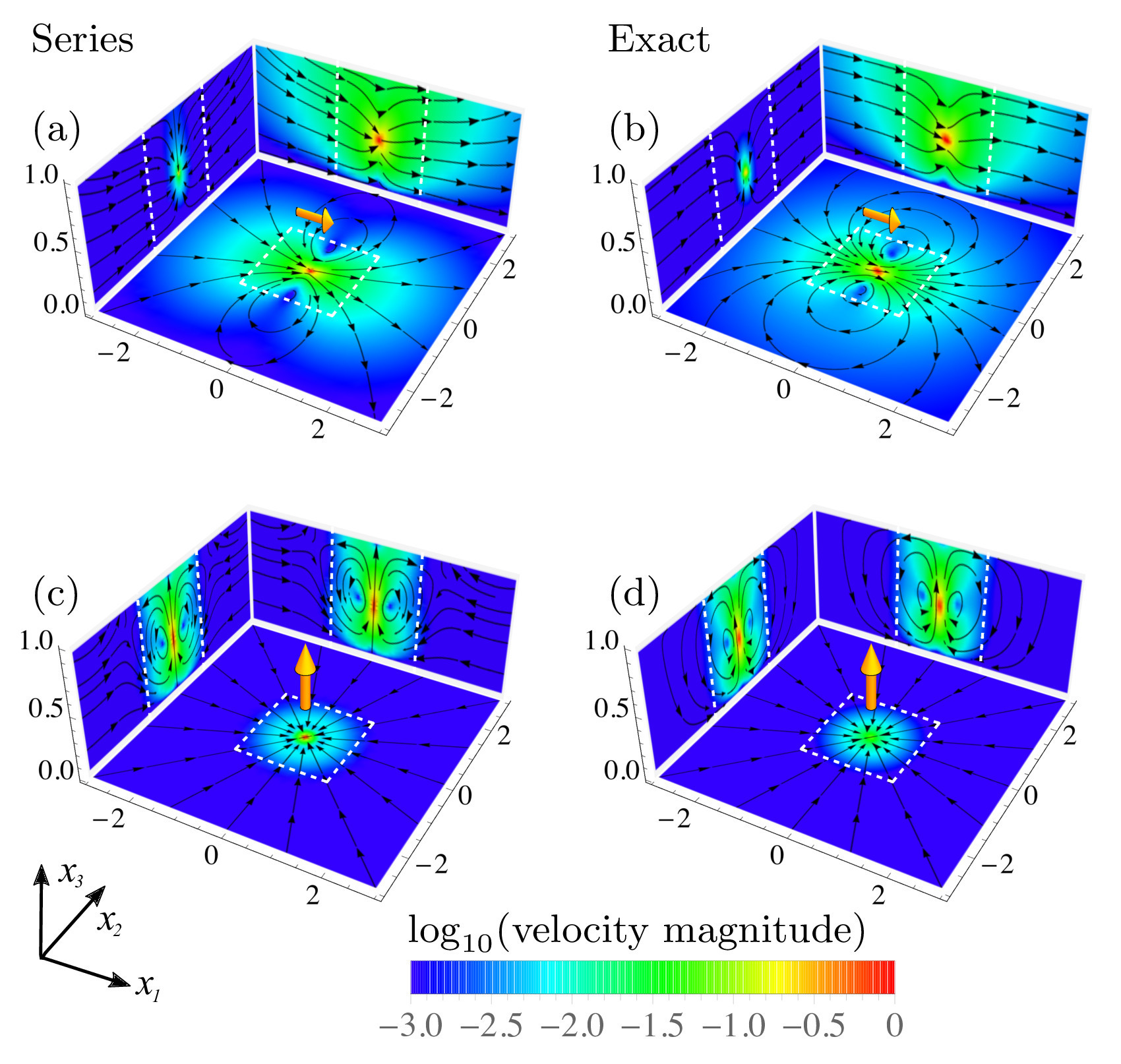} 
\end{center}
\caption{
Comparison of flow fields produced by a Stokeslet in a liquid film using the solution obtained with the recursive series method (left panels; \sect{sec:filmStokeslet}), where $n=9$ images have been used; and the exact solution (right panels; Appendix \sect{appsec:StokesletFlowFilm}).
The film has height $H$, the Stokeslet is located at $h= 0.45H$ just beneath the center, an arbitrary non-trivial point, and it is oriented in the $x_1$ direction (a,b) or in the $x_3$ direction (c,d). 
The planar cuts and colour legends are defined as in \fig{fig:freeSwimmerFlowFields}.
The white dashed lines mark the region $\frac{r_1}{H}, \frac{r_2}{H} < 1$, inside which the series solution is accurate.
}
  \label{fig:freeStokesletFlowFields}  
\end{figure}

\begin{figure}
	\begin{center}
    	\includegraphics[width=\linewidth]{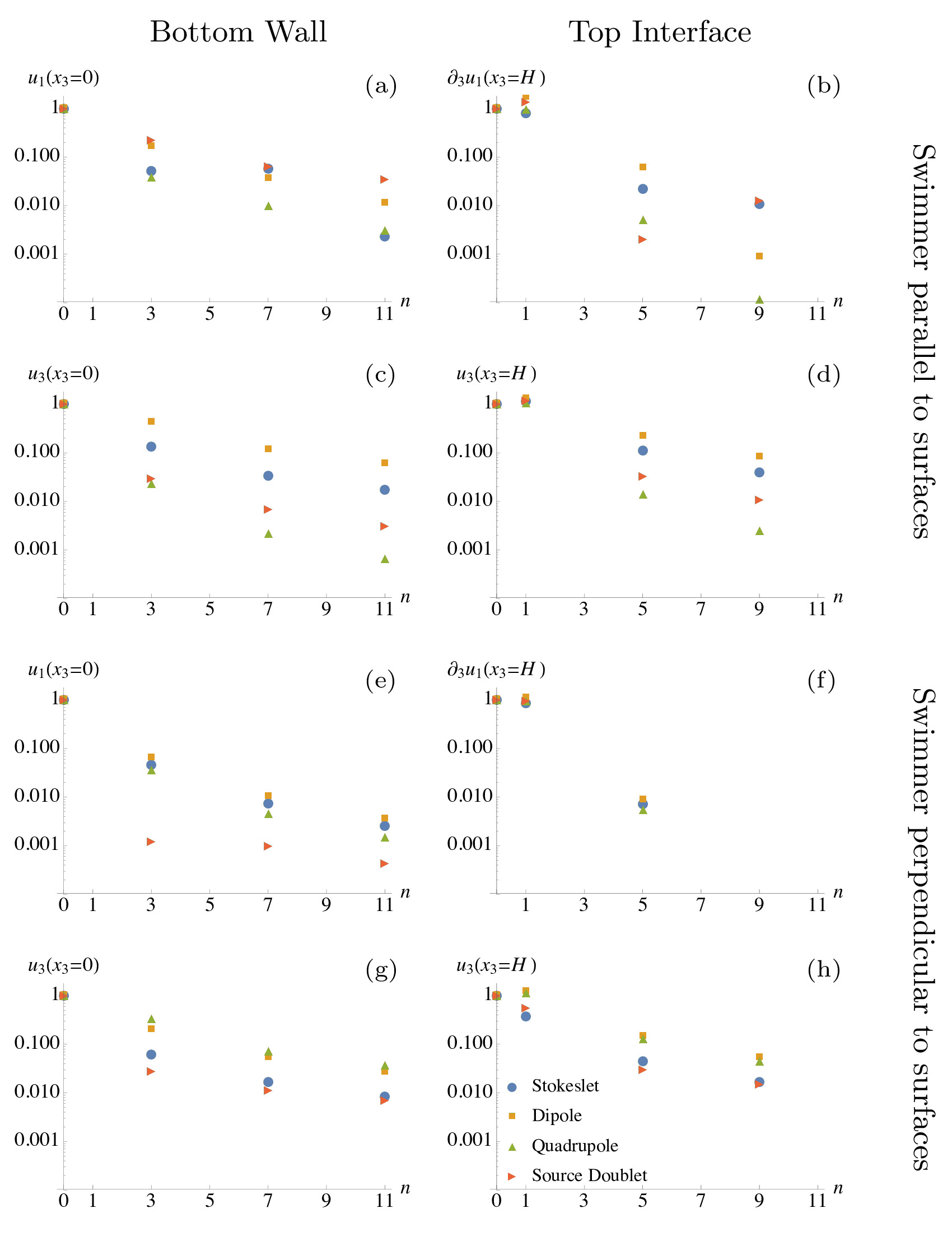}
	\end{center}
  	\caption{
	Convergence of the flow fields generated by a micro-swimmer in a film as a function of the number of images $n$, normalised with respect to the value if only the swimmer itself is present ($n=0$), on a logarithmic scale.
	Only odd values of $n$ are shown, which corresponds to pairs of image systems being added on both sides of the film.
	The swimmer is located just underneath the middle of the film $\vec{y}=(0,0,0.45 H)$, and the flow is sampled at arbitrary points on the two surfaces $\vec{x}=(0,H/4,0)$ and  $\vec{x}=(0,H/4,H)$.
	In panels (a-d) the swimmer is oriented in the $\vec{\hat{e}}_1$ direction, and in panels (e-h) the swimmer is oriented in the $\vec{\hat{e}}_3$ direction.
	The convergence of the boundary conditions is shown at the bottom wall (a,c,e,g),
	and top interface (b,d,f,h) for the normal and parallel flow components. 
	Note that if no marker is shown, the value is exactly zero.
	}
  	\label{appfig:convergence}
	\label{fig:convergence}
\end{figure}

To establish the length scales for which the recursive series solution (\sect{sec:filmStokeslet}) is applicable, we compare it to the exact solution (Appendix \sect{appsec:StokesletFlowFilm}).
There are four important length scales in the system of a micro-swimmer in a liquid film: the organism size $a$, the film height $H$, the distance to the nearest surface $\hat{h} = \min(h,H-h)$, and the size of the flow region of interest, measured by the distance from the swimmer to where the flow is evaluated $r = |\vec{x} - \vec{y}|$.

\fig{fig:freeStokesletFlowFields} shows the flow fields generated by a Stokeslet in a liquid film, using the recursive image series (left panels) and the exact solution (right panels).
The Stokeslet is located at a distance $h=0.45H$ from the bottom wall and is oriented in the direction parallel (top panels) and perpendicular (bottom panels) to the film surfaces.
Close to the point force, where $a \sim r \ll \hat{h} < H$, the flow in a bulk fluid given by the Oseen result (\eq{eq:OseenTensor}) is recovered.
If the film height is increased, with $h$ kept constant such that $a \sim r \sim h \ll H$, then the Blake result (\eq{eq:blakeTensor1}) is recovered (see details in Appendix \sect{appsec:Comparison}, \fig{appfig:LironPlots1}). 
Similarly, if the Stokeslet is located close to the top interface, the local flows can also be described with a single image system (\eq{eq:freeBoundaryTensor}).
However, for a Stokeslet in the middle of the film ($\hat{h} \sim H$), the image systems above and below the film contribute with equal significance.
If one is interested in distant flows, ($a < \hat{h} \sim H \ll r$; \fig{fig:freeStokesletFlowFields}, outside the white dashed lines), then all images are approximately equidistant from the point $\vec{x}$ so that many terms are required in the series to eliminate differences between the two methods.
This regime is equivalent to the thin-film limit (Appendix \sect{appsec:FarField}).
If one is interested in local flows ($a < r < \hat{h} \sim H$; \fig{fig:freeStokesletFlowFields}, inside the white dashed lines), then there is good agreement between the recursive image series (left panels) and the exact solution (right panels), even with a limited number of images ($n=9$). 
This can be understood because in this domain the series converges and can be truncated, as discussed quantitatively below.
Since this work is concerned with the effect of hydrodynamic interactions with film surfaces on the dynamics of micro-swimmers themselves, the recursive image series method is seen to be accurate.

Convergence of the swimmer-generated flow fields at the surfaces is shown in \fig{fig:convergence} as a function of the number of images used in the recursive series.
The values of the flow (or shear rate) at the boundaries are shown, and these must approach zero to accurately calculate the flows.
Here $n=0$ means no images are present, and only the flow generated by the Stokeslet in bulk is considered. 
As $n$ increases more image reflections are included, where the image numberings are defined in \tbl{tab:images}. 
The swimmer location is chosen near the centre of the film, as a worst-case scenario, and we show the flows at the surfaces where the boundary effects are strongest and the convergence is the slowest, with $\hat{h} \sim H \sim r$.
Both at the bottom wall (\fig{fig:convergence}; left panels) and the top interface (\fig{fig:convergence}; right panels), however, the correction to the flow field is small after only a few images. 
Furthermore, with $n = 1,5,9,\dots$ image systems included the boundary conditions at the bottom wall are satisfied exactly, and similarly with $n = 3,7,11,\dots$ the top interface boundary conditions are fully satisfied. 
This feature of the recursive series method can be leveraged if an exact boundary condition is required at one of the two surfaces.
For thin films compared to the swimmer size, or distances much greater than the film height, more images will be required for a given accuracy. 
The convergence can be justified by noting that each term of each image decays as $\sim (-1)^m \left|2mH\right|^{-1}$. 
Since at every image reflection point ($\vec{y}^{(m)}, \vec{Y}^{(m)}$) the leading term is a Stokeslet pointing in the direction opposite to the one at the previous reflection point, this infinite series of alternatingly opposing Stokeslets converges and can be expressed as converging integral expressions Appendix \sect{appsubsec:ImageSeries}. 
For a dipole swimmer the leading terms decay as $\sim (-1)^m \left|2mH\right|^{-2}$, so the infinite series of alternatingly opposing dipoles converges more rapidly.

In short, the recursive series can compute flows accurately in the region $r<H$ for any $h, x_3$ (\fig{fig:freeStokesletFlowFields}, inside the white dashed lines).
This is a region of particular interest since it is, by construction, where the swimmer resides and where perturbations to the flow fields are most significant. 
Secondly, one requires $a \ll \hat{h}$ to evaluate hydrodynamic interactions with surfaces using the multipole expansion, but it is noteworthy that \citet{spagnolie2012hydrodynamics} argue that far-field hydrodynamic interactions give surprisingly accurate results even for small swimmer-wall separations $a \sim \hat{h}$.
When these two conditions are satisfied, we find that $n=9$ images are sufficient to describe micro-swimmer flows with an error less than $\sim 1\%$.
Therefore in this work, we utilise $n=9$ images in all presented figures.

The advantage of the traditional Fourier transform solution is that it is exact and provides access to all regions of the film. 
In particular, it converges rapidly in regions far from the swimmer, $r \gg H$, and therefore a tractable expression for the flows in the far-field limit can be extracted (Appendix \sect{appsec:FarField}). 
On the other hand, this exact solution can be more tedious to handle, especially when taking derivatives, as is necessary for a multipole expansion of a swimming microbe.
In comparison, the image series in \eq{eq:fundamentalTensorInFilm} is constructed purely from the Oseen tensor and its derivatives, and can be manipulated with ease, both analytically and computationally. 
The best choice of method therefore depends on the purpose in mind.
In the far field the Fourier transform method excels, and in the near field the recursive series method is more convenient.  
Since we are interested in local hydrodynamics, we work with the latter in the remainder of this paper.

\section{Swimmer dynamics in a liquid film}
\label{sec:swimmerDynamics}

Knowing the flow fields that a motile microbe produces within a film allows us to model the hydrodynamic interactions with the bounding surfaces. 
In this section, we will describe the effects of these hydrodynamic interactions on the swimmer dynamics.
Because we focus on the effect of surface accumulation at the bottom wall and the liquid air interface, we consider swimmers much smaller than the film height, $a \ll H$. 
Furthermore, for simplicity we do not include swimmer-swimmer interactions.
In this regime, all flows of interest are in the region $r < H$ where the use of the recursive series solution (\eq{eq:fundamentalTensorInFilm}) is appropriate, as discussed in \sect{sec:Comparison}.
 
Cells with a swimming velocity $\vec{v}^\textmds{SW}=v^\textmds{SW}\vec{p}$ in a quiescent film obey the equations of motion
\begin{eqnarray}
 \label{eq:EquationsOfMotion}
 \dot{\vec{y}} &=& \vec{v}^\textmds{SW} + \vec{v}^\textmds{HI} + \vec{v}^\textmds{ST},
 \\
\label{eq:EquationsOfMotion2}
\dot{\vec{p}} &=& \left( \vec{\Omega}^\textmds{HI} + \vec{\Omega}^\textmds{ST} \right) \times \vec{p}, 
\end{eqnarray}
where the swimmer position and orientation are $\vec{y}$ and $\vec{p}$ respectively with $h = y_3$, hydrodynamic interactions with the surfaces are $\vec{v}^\textmds{HI}, \vec{\Omega}^\textmds{HI}$, 
and $\vec{v}^\textmds{ST}, \vec{\Omega}^\textmds{ST}$ account for steric interactions with the surfaces which, for simplicity, are assumed to be adequately represented by hard-sphere interactions \citep{zottl2012nonlinear}.
Additional terms such as background flows, run-tumble dynamics and thermal noise are neglected here, though they can play important roles in real, biological systems. 
For convenience of computation we express the swimmer orientation in spherical polar co-ordinates, $\vec{p} = -(\cos \theta \cos \phi, \sin \theta, \cos \theta \sin \phi)$, and without loss of generality we set $\theta = 0$ at the initial time.
 
The swimmer's motion is modified by the flow field it generates (\eq{eq:swimmerVelocityDef}) because it is advected and rotated by the reflection of this flow in the boundaries. 
This reflected (auxiliary) flow field is $\vec{u}^* = \vec{u} - \vec{u}^{\infty}$, where $\vec{u}^{\infty}$ is the swimmer generated flow in the absence of boundaries.
The surface-induced translational and rotational velocities are then found by solving the Fax\'{e}n relations~ \citep{kimmicrohydrodynamics} for the force-free and torque-free swimmer. Writing terms up to second order in particle length gives
\begin{eqnarray}
\label{eq:FaxenRelations}
\label{eq:FaxenTranslationalVelocity}
\vec{v}^\textmds{HI}(\vec{y}, \vec{p}) &=&
\left( 1+ \tfrac{1}{6} a^2 \gamma^{-2} \nabla^2 \right)
\vec{u}^*(\vec{x})|_{\vec{x} = \vec{y}},
\\
\label{eq:FaxenAngularVelocity}
\vec{\Omega}^\textmds{HI}(\vec{y}, \vec{p}) &=&
\left[ 
\tfrac{1}{2} \bnabla \times \vec{u}^* 
+G \vec{p} \times ( \vec{\Gamma}^*  \cdot \vec{p} ) \right]_{\vec{x} = \vec{y}}, \qquad
\end{eqnarray}
where the derivatives are with respect to the position $\vec{x}$, the geometry factor $G = \tfrac{\gamma^2 - 1}{\gamma^2 + 1} \in [0,1)$ is a function of the aspect ratio $\gamma$ of the elongated swimmer, and ${\boldsymbol\Gamma}^*=(\nabla{\bf u}^*+(\nabla{\bf u}^*)^{T})/2$ is the strain rate tensor.
 
In the following sections, we will describe the effects of individual multipole contributions of the swimmer-generated flow field  (\eq{eq:swimmerVelocityDef}) on the swimmer dynamics within a film \sect{subsec:dipoleHI} - \sect{subsec:rotletDoubletHI}. 


\begin{table*}
\begin{footnotesize}
\begin{tabular}{| L{0.7cm}| L{0.9cm} || C{3cm} |   C{3cm} ||   C{3cm} |  C{0.7cm} |}
\hline
(n)	& Pos.		& $v^\textmds{D}_1$, $v^\textmds{D}_2$ & $v^\textmds{D}_3$ & $\Omega^\textmds{D}_1$, $\Omega^\textmds{D}_2$ & $\Omega^\textmds{D}_3$ \\[5pt] 
\hline 
(1) 	& $\vec{Y}^{(0)}$	& $\frac{1}{\zeta^2}$ & $ -\frac{1}{\zeta^2}$ & $\frac{1}{\zeta^3}$ & 0 \\ 
(2) 	& $\vec{Y}^{(-1)}$	& $0$ & $\frac{2}{3 \left(\zeta-1\right){}^2}$ & $-\frac{1}{\left(\zeta-1\right){}^3}$ & 0 \\ 
(3) 	& $\vec{y}^{(-1)}$	& $-\left(\zeta-2\right) \zeta$ & $\frac{1}{3} \left(-3 \zeta^2+8 \zeta-2\right)$ & $1-2 \zeta$ & 0 \\ 
(4) 	& $\vec{y}^{(1)}$	 	& $\left(\zeta-2\right) \zeta$ & $\zeta^2-\frac{4 \zeta}{3}-\frac{2}{3}$ & $2 \zeta-3$ & 0 \\ 
(5) 	& $\vec{Y}^{(1)}$	& $\frac{4 \left(\zeta-1\right) \zeta}{\left(\zeta+1\right){}^5}$ & $-\frac{2 \left(\zeta^3+13 \zeta^2-\zeta-1\right)}{3 \left(\zeta+1\right){}^5}$ & $\frac{\zeta \left(\zeta+14\right)-3}{\left(\zeta+1\right){}^5}$ & 0 \\ 
(6) 	& $\vec{Y}^{(-2)}$	& $-\frac{1}{\left(\zeta-2\right){}^2}$ & $\frac{1}{\left(\zeta-2\right){}^2}$ & $-\frac{1}{\left(\zeta-2\right){}^3}$ & 0 \\ 
(7) 	& $\vec{y}^{(-2)}$	& $-\frac{1}{4} \left(\zeta-1\right){}^2$ & $\frac{1}{4} \left(-\zeta^2+3 \zeta-1\right)$ & $\frac{1}{8} \left(1-4 \zeta\right)$ &0  \\ 
(8) 	& $\vec{y}^{(2)}$		& $\frac{1}{4} \left(\zeta-1\right){}^2$ & $\frac{1}{4} \left(\zeta^2-\zeta-1\right)$ & $\frac{1}{8} \left(4 \zeta-7\right)$ & 0 \\ 
(9) 	& $\vec{Y}^{(2)}$	& $\frac{\zeta^4+8 \zeta^3+72 \zeta^2-32 \zeta+16}{\left(\zeta+2\right){}^6}$ & $-\frac{\zeta^4+8 \zeta^3+96 \zeta^2-16}{\left(\zeta+2\right){}^6}$ & $\frac{\zeta \left(\zeta \left(\zeta+6\right)+140\right)-56}{\left(\zeta+2\right){}^6}$ &0  \\ 
(10) 	& $\vec{Y}^{(-3)}$	& $\frac{4 \left(\zeta-2\right) \left(\zeta-1\right)}{\left(\zeta-3\right){}^5}$ & $\frac{2 \left(\zeta^3-19 \zeta^2+63 \zeta-57\right)}{3 \left(\zeta-3\right){}^5}$ & $-\frac{\left(\zeta-18\right) \zeta+29}{\left(\zeta-3\right){}^5}$ & 0 \\ 
~ \vdots & ~ \vdots            &  ~ \vdots       &  ~ \vdots        & ~ \vdots       & \vdots     \\  
\hline
\multicolumn{2}{| c || }{prefactor}	& $\frac{3\kappa}{4H^2} p_1 p_3$, $\frac{3\kappa}{4H^2} p_2 p_3$ &	$ \frac{3\kappa}{8H^2} (p_1^2+p_2^2-2p_3^2) $ & $\frac{3 \kappa}{8 H^3} p_2 p_3$, $-\frac{3 \kappa}{8 H^3} p_1 p_3$ & \\  \hline
\end{tabular}
\end{footnotesize}
\caption{Dipolar hydrodynamic interactions of a micro-swimmer with the surfaces of a film. Given are the boundary-induced translational and rotational velocities due to the first image systems (\tbl{tab:images} and \tbl{tab:imagesExplicit}) as a function of non-dimensionalised swimmer position $\zeta = y_3/H$ and orientation $\vec{p} = (p_1,p_2,p_3)$ with respect to the no-slip wall at $\zeta=0$ and free-slip interface at $\zeta=1$. In each case, the functional term listed is to be multiplied by the stated prefactor. For simplicitly, we consider small swimmers with respect to the film height, so the higher-order Fax\'{e}n term proportional to $a^2$ in \eq{eq:FaxenRelations} is omitted here.}
\label{tab:imagesDipoleHI}
\end{table*}

\begin{table*}
\begin{footnotesize}
\begin{tabular}{| L{0.7cm}| L{0.9cm} || C{3cm} |  C{3cm} ||  C{3cm} |  C{0.7cm} |}
\hline
(n)	& Pos.		& $v^\textmds{SD}_1$, $v^\textmds{SD}_2$ & $v^\textmds{SD}_3$ & $\Omega^\textmds{SD}_1$,$\Omega^\textmds{SD}_2$ & $\Omega^\textmds{SD}_3$ 
\\[5pt] 
\hline
(1) 	& $\vec{Y}^{(0)}$	& $\frac{1}{\zeta^3}$ & $\frac{1}{\zeta^3}$ & $\frac{1}{\zeta^4}$ & 0 \\ 
(2) 	& $\vec{Y}^{(-1)}$	& $-\frac{1}{2 \left(\zeta-1\right){}^3}$ & $-\frac{1}{4 \left(\zeta-1\right){}^3}$ & $0$ & 0 \\ 
(3) 	& $\vec{y}^{(-1)}$	& $\frac{1}{2} \left(5-3 \zeta\right)$ & $\frac{1}{4} \left(3 \zeta-7\right)$ & $-1$ & 0 \\ 
(4) 	& $\vec{y}^{(1)}$	 	& $\frac{1}{2} \left(3 \zeta-1\right)$ & $\frac{1}{4} \left(-3 \zeta-1\right)$ & $1$ & 0 \\  
(5) 	& $\vec{Y}^{(1)}$	& $\frac{\zeta \left(\zeta+20\right)-5}{2 \left(\zeta+1\right){}^5}$ & $\frac{\zeta \left(\zeta+32\right)+7}{4 \left(\zeta+1\right){}^5}$ & $\frac{8}{\left(\zeta+1\right){}^5}$ & 0 \\ 
(6) 	& $\vec{Y}^{(-2)}$	& $-\frac{1}{\left(\zeta-2\right){}^3}$  & $-\frac{1}{\left(\zeta-2\right){}^3}$ & $-\frac{1}{\left(\zeta-2\right){}^4}$ & 0 \\ 
(7) 	& $\vec{y}^{(-2)}$	& $\frac{1}{8} \left(5-3 \zeta\right)$ & $\frac{1}{16} \left(3 \zeta-8\right)$ & $-\frac{1}{4}$ &0  \\ 
(8) 	& $\vec{y}^{(2)}$		& $\frac{1}{8} \left(3 \zeta-1\right)$ & $\frac{1}{16} \left(-3 \zeta-2\right)$ & $\frac{1}{4}$ & 0 \\ 
(9) 	& $\vec{Y}^{(2)}$	& $\frac{\zeta \left(\zeta \left(\zeta+6\right)+108\right)-40}{\left(\zeta+2\right){}^6}$ & $\frac{\zeta \left(\zeta \left(\zeta+6\right)+84\right)+32}{\left(\zeta+2\right){}^6}$ & $\frac{\left(\zeta+2\right){}^2+80}{\left(\zeta+2\right){}^6}$ &0  \\ 
(10) 	& $\vec{Y}^{(-3)}$	& $-\frac{\left(\zeta-24\right) \zeta+39}{2 \left(\zeta-3\right){}^5}$ & $-\frac{\left(\zeta-36\right) \zeta+75}{4 \left(\zeta-3\right){}^5}$ & $\frac{8}{\left(\zeta-3\right){}^5}$ & 0 \\ 
~ \vdots & ~ \vdots            &  ~ \vdots       &  ~ \vdots        & ~ \vdots       & \vdots     \\ 
\hline
\multicolumn{2}{| c || }{prefactor}	
& $-\frac{\sigma}{4 H^3} p_1$, $-\frac{\sigma}{4 H^3} p_2$  
&	$-\frac{\sigma}{H^3} p_3$	
&	$\frac{3\sigma}{8H^4} p_2$, $-\frac{3\sigma}{8H^4} p_1$	
&  
\\ 
\hline
\end{tabular}
\end{footnotesize}
\caption{Same as \tbl{tab:imagesDipoleHI}, but for source dipolar hydrodynamic interactions.}
\label{tab:imagesSourceDipoleHI}
\end{table*}
\begin{table*}
\begin{footnotesize}
\begin{tabular}{| L{0.7cm}| L{0.9cm} || C{3cm} |  C{0.7cm} ||  C{3cm} |  C{3cm} |}
\hline
(n)	& Pos.		& $v^\textmds{RD}_1$, $v^\textmds{RD}_2$ & $v^\textmds{RD}_3$ & $\Omega^\textmds{RD}_1$, $\Omega^\textmds{RD}_2$ & $\Omega^\textmds{RD}_3$ 
\\[5pt]  
\hline
(1) 	& $\vec{Y}^{(0)}$	& $0$ & $0$ 		& $-\frac{3}{\zeta^4}$ & $\frac{1}{\zeta^4}$ \\
(2) 	& $\vec{Y}^{(-1)}$	& $-\frac{1}{\left(\zeta-1\right){}^3}$ & $0$ 		& $-\frac{1}{\left(\zeta-1\right){}^4}$ & $\frac{1}{\left(\zeta-1\right){}^4}$ \\
(3) 	& $\vec{y}^{(-1)}$	& $1-\zeta$ & $0$ 		& $3$ & $1$ \\
(4) 	& $\vec{y}^{(1)}$	 	& $\zeta-1$ & $0$ 		& $-3$ & $-1$ \\
(5) 	& $\vec{Y}^{(1)}$	& $\frac{\zeta \left(\zeta+8\right)-1}{\left(\zeta+1\right){}^5}$ & $0$ 		& $\frac{\zeta-15}{\left(\zeta+1\right){}^5}$ & $-\frac{1}{\left(\zeta+1\right){}^4}$ \\
(6) 	& $\vec{Y}^{(-2)}$	& $0$ & $0$ 		& $\frac{3}{\left(\zeta-2\right){}^4}$ & $-\frac{1}{\left(\zeta-2\right){}^4}$ \\
(7) 	& $\vec{y}^{(-2)}$	& $\frac{1}{4} \left(2-\zeta\right)$ & $0$ 		& $\frac{7}{16}$ & $-\frac{1}{16}$ \\
(8) 	& $\vec{y}^{(2)}$		& $\frac{\zeta}{4}$ & $0$ 		& $-\frac{7}{16}$ & $\frac{1}{16}$ \\
(9) 	& $\vec{Y}^{(2)}$	& $\frac{32 \left(2 \zeta-1\right)}{\left(\zeta+2\right){}^6}$ & $0$  & $\frac{-3 \zeta \left(\zeta+4\right)-172}{\left(\zeta+2\right){}^6}$ & $\frac{1}{\left(\zeta+2\right){}^4}$ \\
(10) 	& $\vec{Y}^{(-3)}$	& $-\frac{\left(\zeta-12\right) \zeta+19}{\left(\zeta-3\right){}^5}$ & $0$ 		& $-\frac{\zeta+13}{\left(\zeta-3\right){}^5}$ & $\frac{1}{\left(\zeta-3\right){}^4}$ \\
\vdots & ~ \vdots     & ~ \vdots  &  ~ \vdots        & ~ \vdots       & \vdots     \\ 
\hline
\multicolumn{2}{| c || }{prefactor}	
& $-\frac{3\tau}{4H^3} p_2 p_3$, $\frac{3\tau}{4H^3} p_1 p_3$	
&		
& $\frac{3\tau}{8 H^4} p_1 p_3$, $\frac{3\tau}{8 H^4} p_2 p_3$ 
& $\frac{3\tau}{16H^4} (p_1^2+p_2^2-2p_3^2 )$  
\\
\hline
\end{tabular}
\end{footnotesize}
\caption{Same as \tbl{tab:imagesDipoleHI}, but for rotlet doublet hydrodynamic interactions.}
\label{tab:imagesRotletDipoleHI}
\end{table*}

\begin{figure*}
\begin{centering}
\includegraphics[width = 0.9\linewidth]{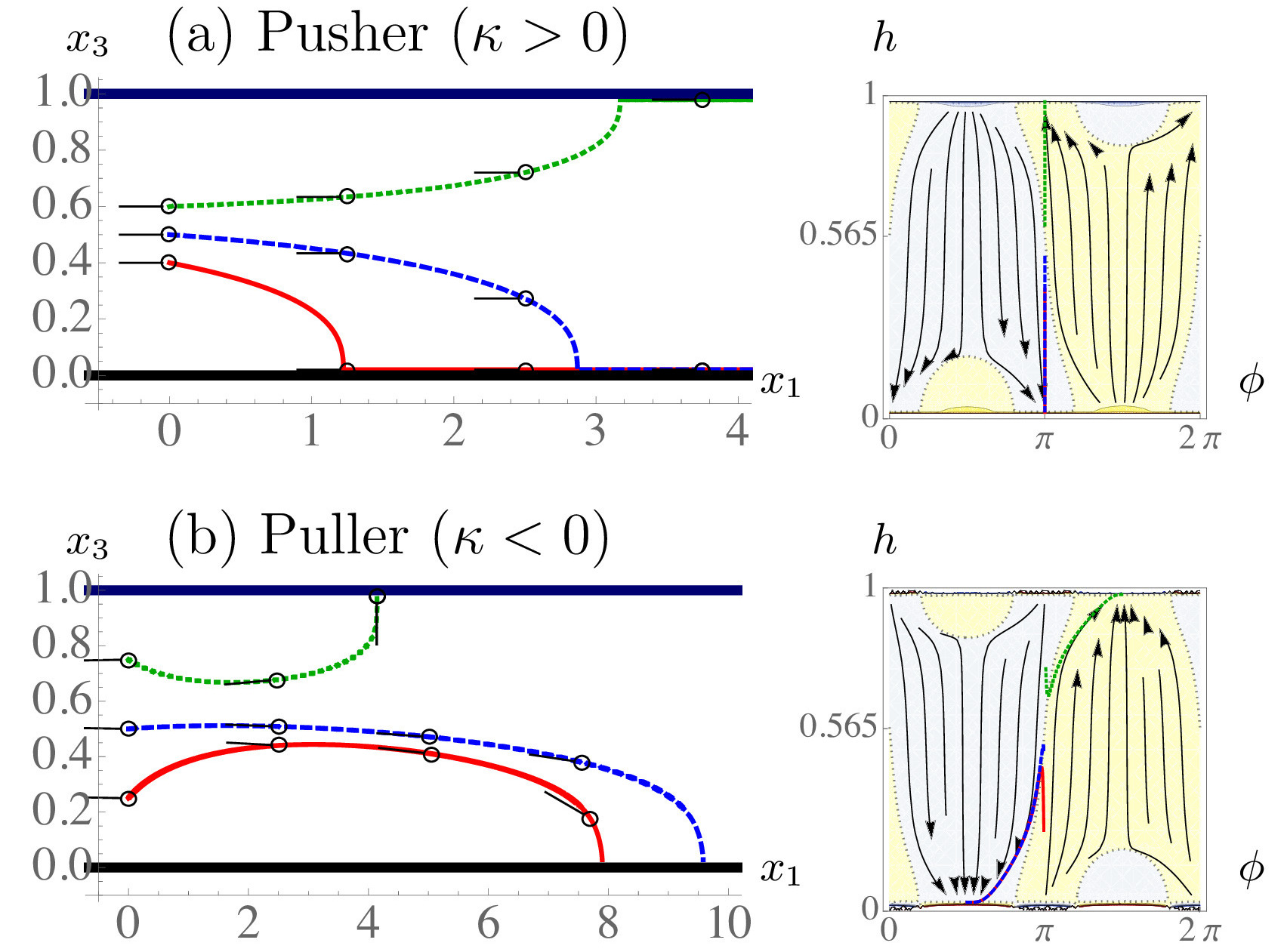}
\caption{The effect of dipolar hydrodynamic interactions on the dynamics of a micro-swimmer in a film.
Swimmer trajectories in real $x_1-x_3$ space are shown in the left hand panels, and the right hand panels show the same dynamics in $\zpos-\phi$ phase space. 
For simplicity we consider spherical swimmer bodies.
 The swimmer size is small compared to the film height, $a =b= H/100$.
The background colours in the phase space plots indicate the sign of the velocity in the $x_3$ direction, where yellow is upward swimming and light blue is downward swimming.
a) A pusher swimmer with $\kappa = 1/20 \left[H^2 v^\textmds{SW} \right]$. 
b) The equivalent puller with $\kappa = -1/20 \left[H^2 v^\textmds{SW} \right]$. 
All other multipole moments are set to zero.
The schematic black micro-organisms indicate the swimming direction.
}
\label{fig:hydrodynamicInteractionPlotsDipole}
\end{centering}
\end{figure*}

\subsection{Dipolar term hydrodynamic interactions}
\label{subsec:dipoleHI}

The primary contribution to the flow field generated by a force-free micro-swimmer at low-Reynolds number is the Stokes dipole. 
A flagellated swimmer with its propulsion mechanism located at the rear of its body is a pusher swimmer with dipole coefficient $\kappa>0$, whereas an organism propelling itself with flagella at the front of its body is modelled as a puller with $\kappa<0$~\citep{lauga2009hydrodynamics}.

To determine the dipolar hydrodynamic interactions with the film surfaces, we insert the auxiliary dipolar flow field (\eq{eq:flowStokesDipole}; $\vec{u}^\textmds{D*} = \vec{u}^\textmds{D} - \vec{u}^{\textmds{D}\infty}$) into \textbf{Eqs. (\ref{eq:FaxenTranslationalVelocity}-\ref{eq:FaxenAngularVelocity})}.
The first two image systems give
\begin{eqnarray}
\label{eq:HI_D_V_X}
v^\textmds{D}_{1} 
&=& \kappa  \frac{\sin (2\phi )}{16} \left(\frac{a^2 \gamma^{-2}}{\left(H-\zpos\right){}^4}-\frac{3 a^2 \gamma^{-2}}{\zpos^4}+\frac{6}{\zpos^2}\right),
\\
\label{eq:HI_D_V_Z}
v^\textmds{D}_{3} 
&=& \kappa \frac{3 \cos (2 \phi )-1}{32} \left(\frac{-a^2 \gamma^{-2}}{\left(H-\zpos\right){}^4}+\frac{3 a^2 \gamma^{-2}}{\zpos^4}+\frac{4}{\left(H-\zpos\right){}^2}-\frac{6}{\zpos^2}\right),
\\
\label{eq:HI_D_O}
\Omega^\textmds{D}_{2} 
&=& \kappa \frac{3 \sin (2\phi )}{64} \left(\frac{4+2G-2 G \cos (2 \phi )}{\left(H-\zpos\right){}^3}+\frac{4+3G -G \cos (2 \phi )}{\zpos^3}\right).
\end{eqnarray}
The complete list of dipolar hydrodynamic interactions due to the first 10 image systems is given in \tbl{tab:imagesDipoleHI} for a small swimmer, with $a \ll H$ in \eq{eq:FaxenTranslationalVelocity}, in terms of the dimensionless position $\zeta=\zpos/H$. 

A pusher-type swimmer is attracted towards both surfaces and tends to align parallel to them (\fig{fig:hydrodynamicInteractionPlotsDipole}a) \citep{berke08, spagnolie2012hydrodynamics}.
The $\zpos-\phi$ phase space in \fig{fig:hydrodynamicInteractionPlotsDipole}a shows stable fixed points at $\zpos=b, H-b;  \phi=0, \pm \pi$ demonstrating that the specific trajectories shown (red, blue and green trajectories) are representative of the general behaviour of pusher swimmers in films. 

A puller, on the other hand, tends to orient perpendicular to the planar boundaries (\fig{fig:hydrodynamicInteractionPlotsDipole}b). 
A puller's stable fixed points are found at $\zpos=a, H-a;  \phi=\pm \pi/2$. 
For both pusher and puller types, the hydrodynamic interactions are a factor of $3/2$ stronger at the solid, no-slip bottom wall than the no-shear free interface. 
All dipolar swimmers initially oriented parallel to the surfaces and located below a critical height $\zpos^*$ will accumulate at the bottom wall. 
The blue trajectories in \fig{fig:hydrodynamicInteractionPlotsDipole}a and \fig{fig:hydrodynamicInteractionPlotsDipole}b correspond to $\zpos=H/2<\zpos^*$ and demonstrate that cells which began swimming parallel to the surfaces along the centre line accumulate at the bottom wall. 
An analytic estimate for this critical height can be obtained using the first two image systems (\eq{eq:HI_D_V_X}-\textbf{\ref{eq:HI_D_O}}).
This gives $\zpos^* \approx \left(3-\sqrt{6}\right)H$ compared to the numerically determined value $\zpos^*=0.565H$, which is different by $2.6\%$. 
In brief, hydrodynamic interactions attract both pushers and pullers more toward the no-slip wall than toward the film interface, when averaged over all initial swimmer positions and orientations. 

\subsection{Source doublet term hydrodynamic interactions}
\label{subsec:sourceDoubletHI}

\begin{figure*}
\begin{centering}
\includegraphics[width = 0.9 \linewidth]{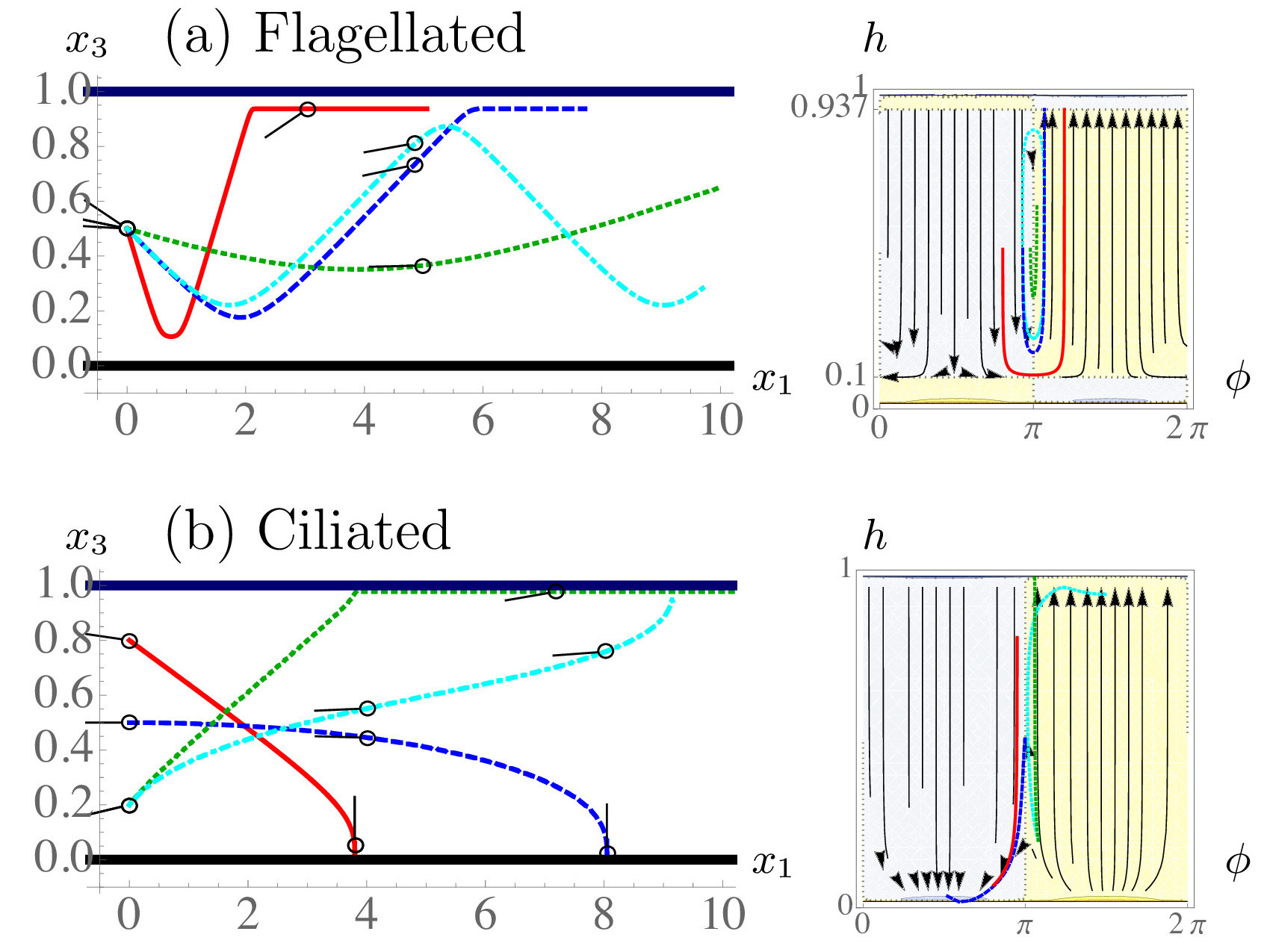}
\caption{
The effect of source doublet hydrodynamic interactions on the dynamics of a micro-swimmer in a film, as defined in \fig{fig:hydrodynamicInteractionPlotsDipole}.
a) A ciliated swimmer with $\sigma = (1/10)^3 \left[H^3 v^\textmds{SW} \right]$ so that the distances of closest approach are $h^\textmds{B}=H/10$ and $H-h^\textmds{T}=0.937H$. 
b) A flagellated sphere with $\sigma = - (1/10)^3 \left[H^3 v^\textmds{SW} \right]$.
The cyan trajectories are for non-spherical swimmers with an aspect ratio of $\gamma=3$ with $a/3 = b = H/100$. 
}
\label{fig:hydrodynamicInteractionPlotsQSD}
\end{centering}
\end{figure*}

For ciliated micro-organisms, such as \textit{Paramecium} and \textit{Volvox}, that rely on local surface deformations as propulsion mechanism, or squirmer swimmers~\citep{blake1971spherical, lauga2009hydrodynamics} with a slip-velocity at their surface such as active colloids \citep{paxton2004catalytic, howse2007self, valadares2010catalytic}, it is expected that the source dipolar coefficient $\sigma >0$. For non-ciliated but flagellated organisms, one expects the coefficient $\sigma < 0$.
\tbl{tab:imagesSourceDipoleHI} lists the hydrodynamic interactions of a small swimmer with the surfaces due to its source doublet flow field. The first two image systems give 
\begin{eqnarray}
\label{eq:HI_QSD_V_X}
v^\textmds{SD}_{1} 
&=& \sigma  \frac{\cos (\phi )}{8} \left(-\frac{2 a^2 \gamma^{-2}}{\zpos^5}+\frac{1}{\left(H-\zpos\right){}^3}+\frac{2}{\zpos^3}\right),
\\
\label{eq:HI_QSD_V_Z}
v^\textmds{SD}_{3} 
&=& \sigma  \frac{\sin (\phi )}{4} \left(-\frac{2 a^2 \gamma^{-2}}{\zpos^5}+\frac{1}{\left(H-\zpos\right){}^3}+\frac{4}{\zpos^3}\right),
\\
\label{eq:HI_QSD_O}
\Omega^\textmds{SD}_{2} 
&=& \sigma  \frac{3\cos (\phi )}{32} \left(\frac{G (\cos (2 \phi )-3)}{\left(H-\zpos\right){}^4}+\frac{-3 G \cos (2 \phi )+9 G+4}{\zpos^4}\right).
\end{eqnarray}

While the force dipole is the most significant contribution to the hydrodynamic interactions attracting swimmers to the film surfaces, we find that the higher-order terms can account for a boundary repulsion if $\sigma >0$ (\fig{fig:hydrodynamicInteractionPlotsQSD}a). 
This can be understood since the source doublet component of the flow field regularises the singular description of the swimmer's flow field and hence gives the swimmer an effective characteristic hydrodynamic size \citep{kimmicrohydrodynamics, mathijssen2015tracer}, $a^\textmds{H} = \sqrt[3]{2\sigma/v^\textmds{SW}}$. 
When the source doublet contribution is significant, the hydrodynamic size can be substantially larger than the characteristic geometric size of the swimmer $a$, which can aid escape from boundaries or keep the swimmers from coming in direct contact with surfaces.
The corresponding hydrodynamic interactions turn a swimmer away from a no-slip wall, so that the distance of closest approach is $h^\textmds{B} = a^\textmds{H}/2^{1/3}$~\citep{mathijssen2015upstream}. 
Likewise, close to a free-slip interface, the closest distance of approach is $h^\textmds{T} = a^\textmds{H} / 2$. 

For spherical ($\gamma =1$) ciliated swimmers the angular velocity vanishes near the top no-shear interface. 
Such swimmers are only turned away from the bottom boundary and therefore the source doublet leads to accumulation at the top surface. 
The red and dark blue trajectories in \fig{fig:hydrodynamicInteractionPlotsQSD}b show this explicitly and the $\zpos-\phi$ phase space shows that all swimmers that initially approach the bottom wall are turned away and accumulate at the film interface where they have the same orientation as that with which they initially approached the bottom wall. 
Ciliated swimmers that initially approach the interface immediately accumulate there, maintaining their initial angle (\fig{fig:hydrodynamicInteractionPlotsQSD}a). 

Elongated ($\gamma >1$) ciliated swimmers, however, are subject to an additional angular velocity at the top surface, which also turns the swimmer away. 
Hence, accumulation at both boundaries can be prevented by the action of the source doublet and the elongated swimmer ($\gamma=3$) oscillates between the two surfaces (cyan trajectory in \fig{fig:hydrodynamicInteractionPlotsQSD}a.) 

Non-ciliated organisms with $\sigma < 0$, conversely, are bound more strongly to the surfaces than ciliated swimmers.
The source doublet hydrodynamics for any swimmer initially oriented towards the wall act to further orient the swimmer towards a head-on collision with the wall (\fig{fig:hydrodynamicInteractionPlotsQSD}b; red and blue trajectories). 
This is because a spherical body that translates past a solid wall also rotates in a rolling fashion due to hydrodynamic interactions~\citep{kimmicrohydrodynamics}.
In the case of a swimmer, this effect points the propulsion direction towards the boundary, thus trapping the swimmer.
The $\zpos-\phi$ phase space shows that all swimmers initially oriented towards the wall approach the stable fixed point at $\zpos=h^\textmds{B};  \phi=\pm \pi/2$. 
Spherical swimmers initially oriented towards the top interface move with a roughly constant $\phi$ and keep that orientation, which is similar to ciliated swimmers (\fig{fig:hydrodynamicInteractionPlotsQSD}a). Elongated swimmers, instead, rotate towards the interface (cyan trajectory).

In summary, higher-order hydrodynamic moments are required to fully characterize swimmer dynamics in confined geometries as large hydrodynamic radii, $a^\textmds{H} \gg a$, can lead to non-trivial trajectories and hence particle distributions.

\subsection{Quadrupolar term hydrodynamic interactions}
\label{subsec:quadrupoleHI}

\begin{figure*}
\begin{centering}
\includegraphics[width = 0.9 \linewidth]{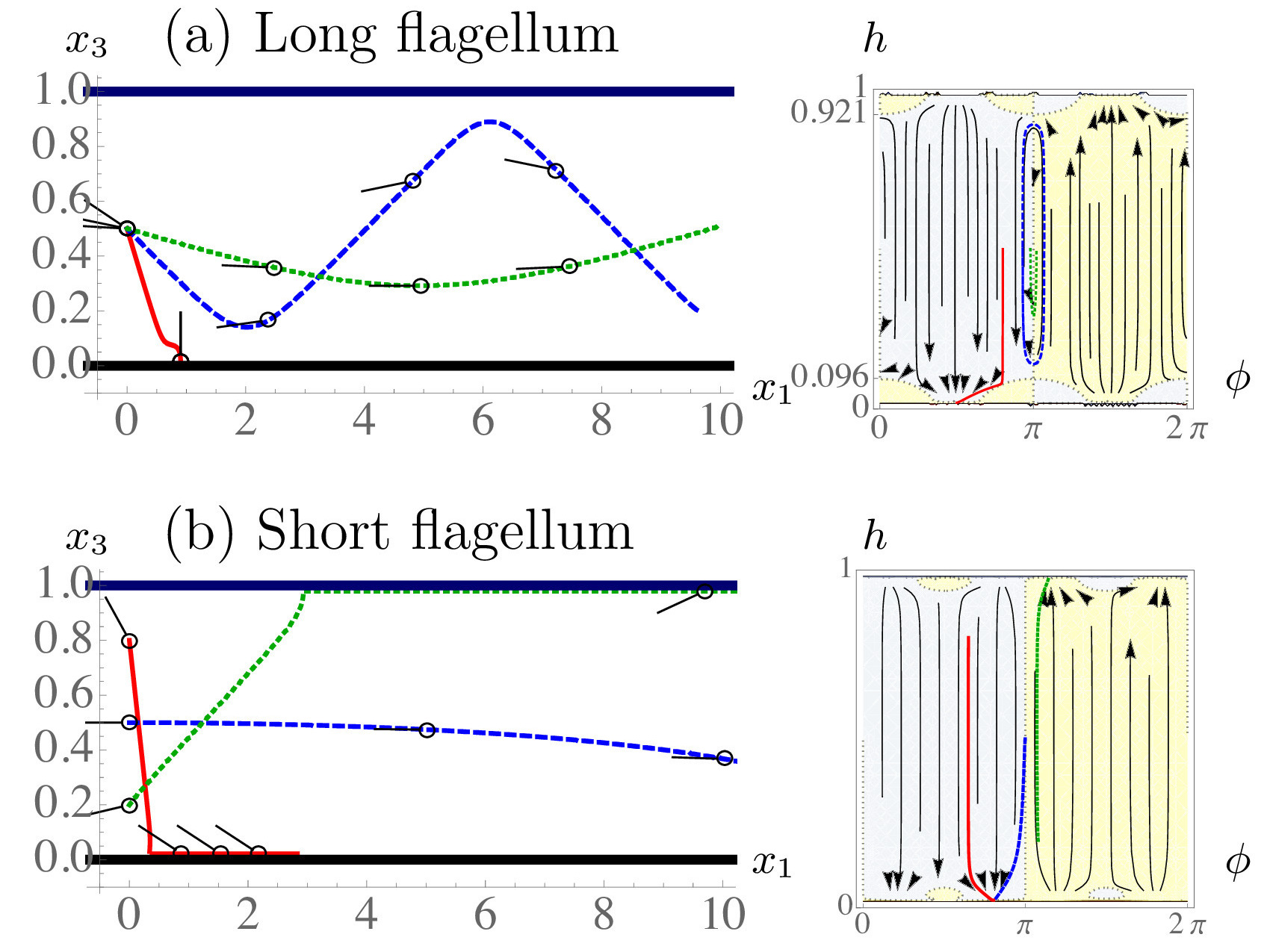}
\caption{The effect of quadrupolar hydrodynamic interactions on the dynamics of a micro-swimmer in a film, as defined in \fig{fig:hydrodynamicInteractionPlotsDipole}.
a) A swimmer with a long flagellum and small body , $\nu = (1/10)^{3} \left[H^3 v^\textmds{SW} \right]$ and $h^\textmds{B}=0.096H$ and $H-h^\textmds{T}=0.921H$.
b) A swimmer with a short flagellum and big body, $\nu = -(1/10)^{3} \left[H^3 v^\textmds{SW} \right]$. 
}
\label{fig:hydrodynamicInteractionPlotsQuadrupole}
\end{centering}
\end{figure*}

Next, we characterise the impact of the quadrupolar term in \eq{eq:swimmerVelocityDef}. This describes the weighting of propulsion forces towards the posterior end of flagellated organisms and therefore accounts for the fore-aft asymmetry. 
Accounting for the first two images gives 
\begin{eqnarray}
\label{eq:HI_Q_V_X}
v^\textmds{Q}_{1} 
&=& \nu  \frac{\cos (\phi )}{128} \Bigg(\frac{3 a^2 \gamma^{-2} (5 \cos (2 \phi )-3)}{\left(\zpos-H\right){}^5} +\frac{4 a^2 \gamma^{-2} (5-11 \cos (2 \phi ))}{\zpos^5}
\nonumber \\ &+&
\frac{4 (3 \cos (2 \phi )-1)}{\left(H-\zpos\right){}^3}+\frac{54 \cos (2 \phi )-26}{\zpos^3}\Bigg),
\\
\label{eq:HI_Q_V_Z}
v^\textmds{Q}_{3} 
&=& \nu  \frac{\sin (\phi )}{32} \Bigg(\frac{a^2 \gamma^{-2} (5 \cos (2 \phi )+1)}{\left(\zpos-H\right){}^5}-\frac{2 a^2 \gamma^{-2} (7 \cos (2 \phi )+3)}{\zpos^5}
\nonumber \\ &+&
\frac{4 (3 \cos (2 \phi )+1)}{\left(H-\zpos\right){}^3}+\frac{2 (9 \cos (2 \phi )+5)}{\zpos^3}\Bigg),
\\
\label{eq:HI_Q_O}
\Omega^\textmds{Q}_{2} 
&=& \nu \frac{3 \cos (\phi )}{512} \Bigg(\frac{2 (3 G \cos (4 \phi )-4 (2 G+5) \cos (2 \phi )-11 G+12)}{\left(H-\zpos\right){}^4}
\nonumber \\ 
&+&\frac{-3 G \cos (4 \phi )+12 (G+4) \cos (2 \phi )+79 G-16}{\zpos^4}\Bigg).
\end{eqnarray}
For swimmers with long flagella and small bodies, the quadrupolar coefficient is positive, $\nu>0$ (\fig{fig:hydrodynamicInteractionPlotsQuadrupole}a). 
For small angles of approach measured from the surface tangent (blue and green trajectories), the swimmer is turned away from the top and bottom surfaces. 
Such swimmers with $\phi\approx\pi$ perpetually oscillate between the two bounding surfaces. 
However, for larger angles of approach the swimmer can get stuck at the fixed point $\phi=\pi/2$ pitched downwards (\fig{fig:hydrodynamicInteractionPlotsQuadrupole}b, red trajectory), or pitched upwards at $\phi=-\pi/2$.
This mechanism could help prevent the flagella from touching the surface, facilitating an easier escape. 
For a spherical swimmer (as in \fig{fig:hydrodynamicInteractionPlotsQuadrupole}) both fixed points are stable. 
However, the stability decreases at the bottom wall with increasing elongation $\gamma$, which favours top-surface accumulation. 

Conversely, swimmers with a large body and small flagella, i.e.\ $\nu<0$, are only weakly rotated towards the boundaries (\fig{fig:hydrodynamicInteractionPlotsQuadrupole}b, blue trajectory). 
Interestingly, such swimmers have different fixed points than the others:
The majority of the fixed points in \textbf{Figs. \ref{fig:hydrodynamicInteractionPlotsDipole}, \ref{fig:hydrodynamicInteractionPlotsQSD}, \ref{fig:hydrodynamicInteractionPlotsQuadrupole}} are at $\phi=0,\pm\pi/2, \pm \pi$ with the exception of \fig{fig:hydrodynamicInteractionPlotsQuadrupole}b where swimmers have fixed points at an exceptional angle to the surfaces ($\phi=\pm \pi/4$). 

Like the source doublet hydrodynamics, the quadrupolar term turns the swimmer away from both the no-slip and no-shear surfaces for small angles of approach measured from the surface tangent. 
The distance of nearest approach allowed by the quadrupolar term is $h^\textmds{B} = \sqrt[3]{7\nu/8 v^\textmds{SW}}$ for the no-slip wall and $h^\textmds{T} = \sqrt[3]{\nu/2 v^\textmds{SW}}$ for the no-shear interface. 
While it is well known that it is energetically favourable for small colloids to reside at interfaces, this is not commonly observed for microbes in the presence of air-water interfaces. 
The source doublet and quadrupolar distance of nearest approach provides a possible explanation for cases when $a < h^\textmds{T}$.

\subsection{Rotlet doublet term hydrodynamic interactions}
\label{subsec:rotletDoubletHI}

\begin{figure}
\begin{center}
\includegraphics[width = 0.9 \linewidth]{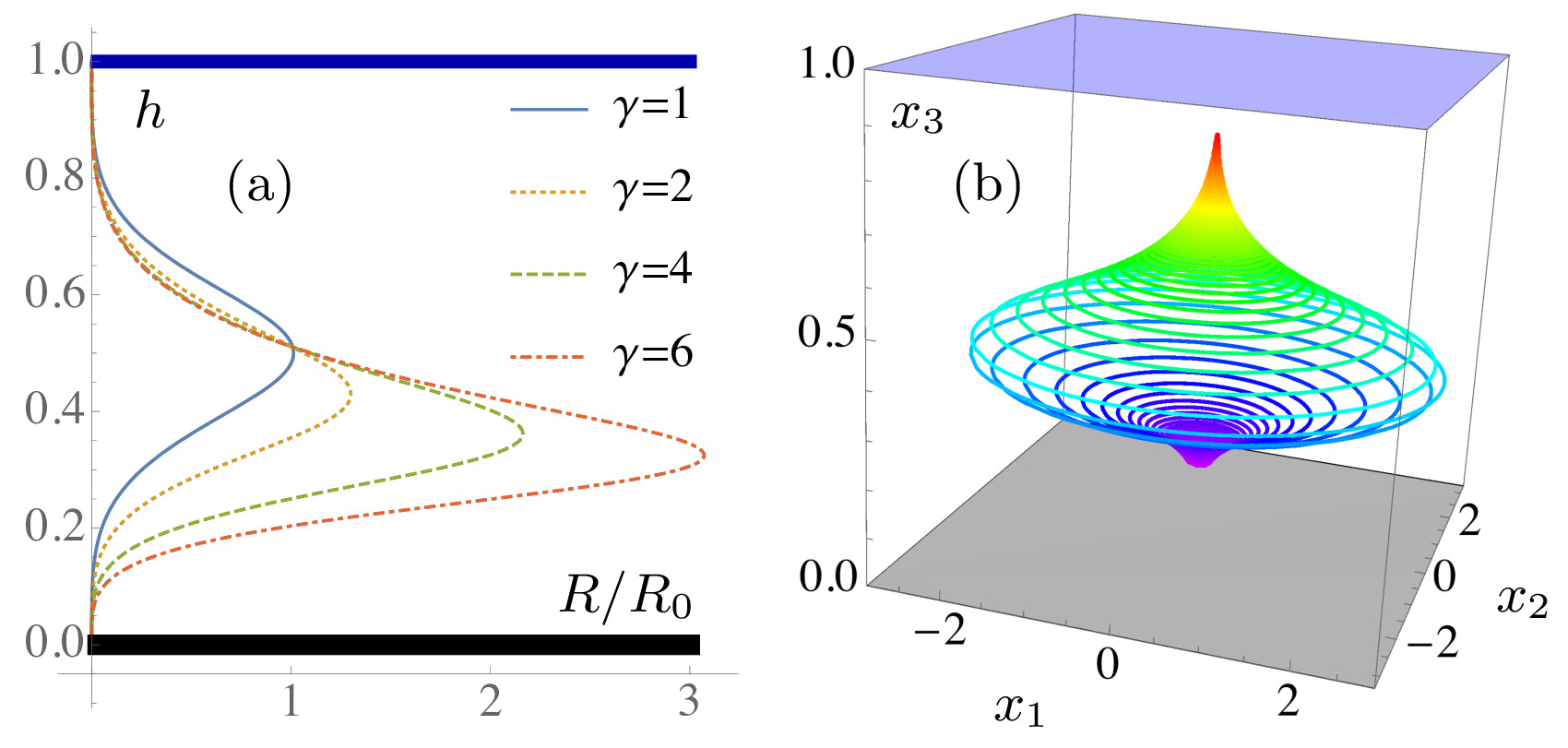}
\end{center}
\caption{
Spiralling trajectories in a film due to the counter-rotation of a swimmer's head and tail. 
(a) Radius of curvature $R(\zpos, \gamma)$ of the swimmer moving in circles in the plane parallel to the surfaces for various aspect ratios $\gamma$, normalised with respect to $R_0 = R(\tfrac{H}{2},1)$.
(b) An example trajectory of an elongated swimmer ($\gamma=4$) that has just escaped from the bottom wall, with a small out-of-plane angle $\phi = 0.1^\circ$. Colours indicate time passing, ranging from violet to red.
The rotlet doublet coefficient is $\tau = 1/10 \left[H^3 v^\textmds{SW}\right]$, all other multipole coefficients are set to zero, and $n=9$ images have been used.
}
\label{fig:RotletHIspiral}
\end{figure}

Lastly, we describe the hydrodynamic interactions due to the rotlet doublet flow field of strength $\tau$, representing the counter-rotating swimmer head and tail. \tbl{tab:imagesRotletDipoleHI} lists the hydrodynamic interactions due to this term in the flow field.
A swimmer immediately above a single no-slip boundary moves in circles in the plane parallel to the boundary. 
\textit{E. coli} bacteria are observed to move in the clockwise direction, as seen from the liquid side, but in the counter-clockwise direction near a free-slip surface \citep{berg1990chemotaxis, frymier1995three, diluzio2005escherichia, spagnolie2012hydrodynamics, leonardo11}.
In a film, these effects are additive.
\fig{fig:RotletHIspiral}(a) shows the radius of curvature of the circular trajectory $R(\zpos, \gamma)$ as a function of the swimmer's position in the film $\zpos$ and its aspect ratio $\gamma$.
Using the first two image systems ($n=2$) gives an analytic estimate of the radius of curvature 
\begin{eqnarray}
\label{eq:filmRadiusOfCurvature}
 R(\zpos,\gamma) &\approx \frac{16 v^\textmds{SW} }{3 \tau} \left(\frac{1- G}{\zpos^4}+\frac{1+G}{(H-\zpos)^4}\right)^{-1} , 
\end{eqnarray}
where $G = \tfrac{\gamma^2 - 1}{\gamma^2 + 1}$. 
This function is maximised in the middle of the film, at $\zpos = H/2$, if the swimmer has a spherically symmetric body. 
The maximum radius of curvature is then $R_0 = R(\tfrac{H}{2},\gamma=1) \approx v^\textmds{SW} H^4/6 \tau$.
For an elongated swimmer, the function $R(\zpos,\gamma>1)$ is no longer symmetric about the film centreline: the radius of curvature is larger near the bottom wall but smaller near the top surface.
Consequently, elongated swimmers' trajectories bend more near the water-air interface (\fig{fig:RotletHIspiral}b). 

In microfluidic experiments, the radius of curvature of swimming trajectories can be used to determine whether micro-organisms are located near the top or bottom microscope slide, as was shown qualitatively in experiments by \citet{guidobaldi2015disrupting}.
More quantitatively, the knowledge of the function $R(\zpos,\gamma)$ now allows one to compute the vertical position $\zpos$ by inverting \eq{eq:filmRadiusOfCurvature}. 
The rotlet doublet coefficient $\tau$ can also be determined with this equation using the maximum in radius of curvature in the middle of the film.
This may be more reliable and effective than earlier calibration techniques used near a single surface, where an accurate measurement of the swimmer-wall separation is critical and higher-order multipole terms and lubrication effects are hard to separate out.

Between two no-slip walls such as two glass slides, the rotary coefficient $\tau$ is challenging to determine because there is no maximum radius of curvature $R_0$.
However, once the parameters $\gamma, \tau, H$ are deduced in the film, the vertical position $\zpos$ between two microscope slides can be computed from experimentally measured radii of curvature, using the equivalent of \eq{eq:filmRadiusOfCurvature} for two paralel no-slip plates: $R(\zpos,\gamma) \approx \frac{16 v^\textmds{SW} }{3 \tau} \left(\tfrac{1- G}{\zpos^4}-\tfrac{1-G}{(H-\zpos)^4}\right)^{-1}$. 
Moreover, the parameters can be used in different experimental geometries, such as a lubrication layer (solid-liquid-liquid) or a floating oil film (liquid-liquid-air) for which similar expressions for $R(\zpos,\gamma)$ can be computed.

 \section{Conclusions}
 \label{sec:conc}

Surfaces and interfaces are ubiquitous in the microscopic world of microbial swimmers and one of the most important mechanisms by which they impact the behaviour and dynamics of microscopic life is through hydrodynamic flows. 
To analyse these, we have derived the Stokes flow Green's function in a viscous film via a recursion relation for a series of images.
The multipole expansion of this fundamental solution provides the detailed flow fields of a force- and torque-free micro-swimmer.
By comparing this recursive series to the exact solution, we find that $n=9$ images are sufficient to describe local hydrodynamic interactions with the bounding surfaces. 

Using these results, we studied surface accumulation of swimmers in a thick film, distinguishing between the wall and the liquid-air interface.
Here the hydrodynamic flows determine the trajectory and dynamics of a single organism.  
Pusher- and puller-type dipolar swimmers are attracted to both boundaries, though there is a long-ranged bias to accumulate at the no-slip wall.
We give an analytic approximation distinguishing which initial conditions lead to accumulation at the solid wall.
However, we also find that quadrupolar and source doublet hydrodynamic interactions are important in modulating the interface versus wall accumulation, regardless of initial conditions. 
Ciliated swimmers (with a positive source doublet moment) are turned away from the bottom wall and accumulate at the the top interface.
On the other hand, non-ciliated swimmers (with a negative source doublet moment) are more strongly bound to the no-slip wall because the swimming direction is rotated down towards the rigid boundary.
Additional quadrupolar terms due to the swimmer's fore-aft asymmetry further contribute to this, but also result in more complicated dynamics such as trajectories oscillating between the surfaces.
When the rotlet doublet contribution is included the swimmer possesses rotary dynamics within the plane and we have analytically predicted the radius of curvature of these trajectories at all points within viscous films for both spherical and elongated swimmers. 
The formula allows one to first calibrate the rotlet doublet coefficient in a film, and then determine the out-of-plane position of swimmers in a film or other geometries, such as between microscope slides, from experimentally available radii of curvature.

In addition, we note that decreasing the film height leads to the appearance of recirculating flow patterns, for which we provide expressions in the thin-film limit.
These flows could potentially affect motility-related traits such as nutrient uptake and the energy expenditure of cells within films. 
Likewise, the screening of hydrodynamic flows within films seen here is expected to have important consequences for mechanosensing and predator-prey interactions between microbes within films.
These flow fields are a necessary prerequisite for future simulations of the dynamics of micro-swimmers and of motility-based phenomena within films.
Particularly, our thin-film limit expressions could be used to model swimmer-swimmer interactions in films of thickness comparable to the organism size, where swarming and collective motion have been observed.

Microscopic environments are littered with boundaries and interfaces that provide microbial life with a variety of potential benefits or detriments. 
Liquid-air interfaces may be the source of essential resources, such as heightened levels of oxygen. 
Likewise, solid surfaces can accrue sediments including nutrients. 
Since surfaces allow the accumulation of extracellular matrices and biofilms, studying individual swimming dynamics is an important contribution to understanding the initialisation of bacterial colonies and fouling in films.
We have shown how hydrodynamics affect these dynamics near these boundaries in a liquid film on a solid substrate, but the same method can be extended straightforwardly to account for different experimental geometries such as a channel, a lubrication layer or a floating oil film, recently of interest because of oil-consuming bacteria at spill sites \citep{Karimi2015,Stebe2015}.
These ideas can also be used to study biological traits, such as species-specific interactions with surfaces, background flows, and run-tumble dynamics of microbial swimmers \citep{mathijssen2015hotspots}.


\section*{Acknowledgements}
This work was supported through funding from the ERC Advanced Grant MiCE (291234 MiCE) and we acknowledge EMBO funding to T.N.S (ALTF181-2013).


\appendix

\section{Exact solution of Stokeslet flow in a liquid film}
\label{appsec:StokesletFlowFilm}

In this section, we derive the flow fields generated by a Stokeslet in a liquid film of uniform thickness using a Fourier transform method. 
We follow the method of \citet{liron1976Stokes} for a Stokeslet between parallel flat plates using the appropriate boundary conditions for a film (no-slip at the bottom wall and no-shear at the top interface). 

\subsection{Formulation of the problem}
\label{appsubsec:ProblemFormulation}

In order to find the flows due to a Stokeslet in the liquid film, we split the solution in two parts. 
For the velocity and pressure Green's functions, respectively, we write
\begin{eqnarray}
\mathcal{F}_{ij} &=& \mathcal{V}_{ij} + \mathcal{W}_{ij}, \qquad i,j \in \{1,2,3\}, \\
\mathcal{P}_{j} &=& \mathcal{Q}_j + \mathcal{S}_j,
\end{eqnarray}
where the first part ($\mathcal{V}_{ij}$, $\mathcal{Q}_j$), which we call the series part, accounts for the singularity in the film and satisfies some of the boundary conditions at the film surfaces.
As detailed in the next section, the series part is composed of an infinite series of image reflections of the Oseen tensor (\eq{eq:OseenTensor}), and contains the Stokeslet singularity in the region $0 \le x_3 \le H$. 

The second part ($\mathcal{W}_{ij}$, $\mathcal{S}_j$), called the auxiliary solution, is then added to the first to satisfy the other boundary conditions.
Because this part does not contain any singularities in the region $0 \le x_3 \le H$ the Stokes equations reduce to the homogeneous equations
\begin{eqnarray}
\label{appeq:StokesEqn2}
\partial_i \mathcal{S}_j &=& \mu \nabla^2 \mathcal{W}_{ij}, \\
\label{appeq:StokesEqnB2}
\partial_i \mathcal{W}_{ij} &=& 0,
\end{eqnarray}
with boundary conditions
\begin{eqnarray}
\label{appeq:BCB1}
\mathcal{W}_{ij} &=& - \mathcal{V}_{ij} \quad \textmd{on $x_3 = 0$}, \\
\label{appeq:BCB2}
\mathcal{W}_{3j} &=& - \mathcal{V}_{3j} \quad \textmd{on $x_3 = H$}, \\
\label{appeq:BCB3}
\partial_3 \mathcal{W}_{\alpha j} &=& - \partial_3 \mathcal{V}_{\alpha j} \quad \textmd{on $x_3 = H$},
\end{eqnarray}
where here and throughout this Appendix, the Greek-letter indices $\alpha, \beta \in \{1,2\}$ only run over the directions parallel to the film surfaces.

\subsection{Infinite series of image reflections}
\label{appsubsec:ImageSeries}

The first part of the solution consists of the infinite series of image reflections of the Stokeslet flow in bulk, as in the work of \citet{liron1976Stokes}.
The images reflected in the top and bottom surface are located at 
\begin{eqnarray}
\vec{y}^{(m)} &=& (y_1,y_2, y_3-2mH), \\
\vec{Y}^{(m)} &=& (y_1,y_2, -y_3-2mH), \quad m = 0, \pm 1, \pm 2, \dots
\end{eqnarray}
Here the original Stokeslet is located at $\vec{y}^{(0)}$, and the relative distances with respect to any point in the film are defined as $\vec{r}^{(m)} = \vec{x} - \vec{y}^{(m)}$ and $\vec{R}^{(m)} = \vec{x} - \vec{Y}^{(m)}$.
Hence, the series part of the total flow field is
\begin{eqnarray}
\label{appeq:ImageSeries}
\mathcal{V}_{ij} = \frac{1}{8 \pi \mu} \sum_{m=-\infty}^{\infty} 
\left[ 
\left( \frac{\delta_{ij}}{r^{(m)}} + \frac{r_{i}^{(m)} r_{j}^{(m)}}{(r^{(m)})^3} \right) - \left( \frac{\delta_{ij}}{R^{(m)}} + \frac{R_{i}^{(m)} R_{j}^{(m)}}{(R^{(m)})^3} \right)
\right],
\end{eqnarray}
where $r^{(m)} = |\vec{r}^{(m)}|$ and $R^{(m)} = |\vec{R}^{(m)}|$, and the corresponding pressure is
\begin{eqnarray}
\label{appeq:ImageSeriesPressure}
\mathcal{Q}_j = \frac{1}{4 \pi} \sum_{m=-\infty}^{\infty} 
\left[ 
\frac{r_{j}^{(m)}}{(r^{(m)})^3} -\frac{R_{j}^{(m)}}{(R^{(m)})^3}
\right].
\end{eqnarray}
This flow and pressure combination satisfies the Stokes equations, as the point force at $\vec{y}^{(0)}$ is included, and all other images lie outside the fluid region $0\le x_3 \le H$.

Note that this image series is \textit{not} the same as the series discussed in \sect{sec:filmStokeslet}. Here we only consider Stokeslet reflections, and not successive \citet{blake1971note} reflections that contain derivatives of the Stokeslet as well.
Therefore, this series can be written as an integral expression without any truncation. 
This is achieved with the Lipshitz integral 
\begin{eqnarray}
\label{appeq:Lipshitz}
\frac{1}{\sqrt{\rho^2+a^2}} = \int_0^\infty J_0 (\rho \lambda) e^{-|a| \lambda} d \lambda
\end{eqnarray}
where we identify $\rho^2 = (x_1-y_1)^2 + (x_2-y_2)^2 = r_1^2 + r_2^2 = r_\alpha r_\alpha$, with the standard summation convention, but with the indices $\alpha, \beta$ taking the values 1 and 2 only. 
Hence, by summing the geometric series of the exponential factors, the velocity field can be written as 
\begin{eqnarray}
\label{appeq:ImageIntegral}
4 \pi \mu \mathcal{V}_{ij} 
&=& \delta_{ij} \int_0^\infty J_0 (\rho \lambda) \frac{\sinh{\lambda h}}{\sinh{\lambda H}} \sinh{\lambda (H-x_3)} d\lambda 
\nonumber \\
&+& \delta_{i\alpha} \delta_{j\beta} \frac{r_\alpha r_\beta}{\rho} \int_0^\infty \lambda J_1 (\rho \lambda) \frac{\sinh{\lambda h}}{\sinh{\lambda H}} \sinh{\lambda (H-x_3)} d\lambda 
\nonumber \\
&-& \delta_{i3} \delta_{j3} \int_0^\infty \lambda J_0 (\rho \lambda) \frac{d}{d\lambda} \left[ \frac{\sinh{\lambda h}}{\sinh{\lambda H}} \sinh{\lambda (H-x_3)}\right] d\lambda 
\nonumber \\
&+& \textmd{sgn}(x_3 - h) (\delta_{i3} \delta_{j \alpha} + \delta_{i \alpha} \delta_{j3}) r_\alpha \int_0^\infty \lambda J_0 (\rho \lambda) \frac{\sinh{\lambda h}}{\sinh{\lambda H}} \cosh{\lambda (H-x_3)} d\lambda
\nonumber \\
&& \textmd{for } x_3 \ge h.
\end{eqnarray}
For $x_3 <h$, \eq{appeq:ImageIntegral} is used but with $x_3$ replaced by $H-x_3$ and $h$ by $H-h$ under the integral signs.
Similarly, the pressure is 
\begin{eqnarray}
\label{appeq:ImageIntegralPressure}
2 \pi \mathcal{Q}_j 
&=& \delta_{j\alpha} \frac{r_\alpha}{\rho} \int_0^\infty \lambda J_1 (\rho \lambda) \frac{\sinh{\lambda h}}{\sinh{\lambda H}} \sinh{\lambda (H-x_3)} d\lambda 
\nonumber \\
&+& \textmd{sgn}(x_3 - h) \delta_{j3} \int_0^\infty \lambda J_0 (\rho \lambda) \frac{\sinh{\lambda h}}{\sinh{\lambda H}} \cosh{\lambda (H-x_3)} d\lambda,
\quad x_3 \ge h.
\end{eqnarray}
Again, for $x_3 <h$, one replaces $x_3$ by $H-x_3$ and $h$ by $H-h$ under the integral signs only, using the minus sign for the second term.

From the infinite series expressions (\ref{appeq:ImageIntegral}--\ref{appeq:ImageIntegralPressure}), we derive the boundary conditions (\ref{appeq:BCB1}--\ref{appeq:BCB3}) on the auxiliary solution $\mathcal{W}_{ij}$. 
On the bottom wall (using \ref{appeq:ImageIntegral} for $x_3<h$), we have
\begin{eqnarray}
\label{appeq:ImageIntegralBCS1}
4 \pi \mu \mathcal{W}_{ij} (x_3 = 0)
&=&  (\delta_{i3} \delta_{j \alpha} + \delta_{i \alpha} \delta_{j3}) r_\alpha \int_0^\infty \lambda J_0 (\rho \lambda) \frac{\sinh{\lambda (H-h)}}{\sinh{\lambda H}} d\lambda.
\end{eqnarray}
Similarly, on the liquid-air interface
\begin{eqnarray}
\label{appeq:ImageIntegralBCS2}
4 \pi \mu \mathcal{W}_{3j} (x_3 = H)
&=& - \delta_{j \alpha} r_\alpha \int_0^\infty \lambda J_0 (\rho \lambda) \frac{\sinh{\lambda h}}{\sinh{\lambda H}} d\lambda 
\qquad \quad \textmd{for } i=3,
\\
\label{appeq:ImageIntegralBCS3}
 \partial_3 4 \pi \mu \mathcal{W}_{ij} (x_3 = H)
&=& \delta_{ij} \int_0^\infty \lambda J_0 (\rho \lambda) \frac{\sinh{\lambda h}}{\sinh{\lambda H}} d\lambda
\nonumber \\
&+& 
\delta_{i\alpha} \delta_{j\beta} \frac{r_\alpha r_\beta}{\rho} \int_0^\infty \lambda^2 J_1 (\rho \lambda) \frac{\sinh{\lambda h}}{\sinh{\lambda H}} d\lambda 
\quad \textmd{for } i=1,2.
\end{eqnarray}
In the next two sections, we transform the equations (\ref{appeq:StokesEqn2}--\ref{appeq:StokesEqnB2}) together with these boundary conditions into Fourier space and hence find a solution.

\subsection{Auxiliary solution in Fourier Space}
\label{appsubsec:AuxiliaryFourier}

To proceed, we define the two-dimensional Fourier transform and its inverse transform for an arbitrary function $\psi(r_1, r_2, x_3)$ as
\begin{eqnarray}
\label{appeq:FourierTransform}
\hat{\psi}(\lambda_1, \lambda_2, x_3) &=& \frac{1}{2\pi} \int_{-\infty}^\infty \int_{-\infty}^\infty \psi(r_1, r_2, x_3) e^{+\ci(\lambda_1 r_1 + \lambda_2 r_2)} dr_1 dr_2,
\\
\label{appeq:InverseTransform}
\psi(r_1, r_2, x_3) &=& \frac{1}{2\pi} \int_{-\infty}^\infty \int_{-\infty}^\infty \hat{\psi}(\lambda_1, \lambda_2, x_3) e^{-\ci(\lambda_1 r_1 + \lambda_2 r_2)} d\lambda_1 d\lambda_2.
\end{eqnarray}
The Stokes equations (\ref{appeq:StokesEqn2}--\ref{appeq:StokesEqnB2}) are then transformed into 
\begin{eqnarray}
\label{appeq:TransformedStokes1}
-\ci \lambda_\alpha \delta_{\alpha i} \hat{\mathcal{S}}_j + \delta_{i3} \frac{\partial}{\partial x_3} \hat{\mathcal{S}}_j &=& \mu \left(\frac{\partial^2}{\partial x_3^2} - \xi^2 \right) \hat{\mathcal{W}}_{ij},
\\
\label{appeq:TransformedStokes2}
-\ci \lambda_\alpha \hat{\mathcal{W}}_{\alpha j} + \frac{\partial}{\partial x_3} \hat{\mathcal{W}}_{3j} &=& 0,
\end{eqnarray}
where the conjugate of $\rho^2$ is $\xi^2 = \lambda_1^2 + \lambda_2^2$.
The pressure must obey the Laplace equation $\nabla^2 \mathcal{S}_j = 0$ and so we have $(\partial_3^2 - \xi^2) \hat{\mathcal{S}}_j = 0$. 

The general solution to this equation, following Liron \& Mochon, can be written as
\begin{eqnarray}
\label{appeq:PressureSolution}
\hat{\mathcal{S}}_j = B_j \sinh{\xi(H-x_3)} + C_j \cosh{\xi(H-x_3)},
\end{eqnarray}
and the solution to equation (\ref{appeq:TransformedStokes1}) is then
\begin{eqnarray}
\label{appeq:VelocitySolution}
\hat{\mathcal{W}}_{ij} 
&=& B_{ij} \sinh{\xi(H-x_3)} + C_{ij} \cosh{\xi(H-x_3)}
\nonumber \\
&+& (B_j \delta_{i3} + C_j \delta_{\alpha i} \ci \lambda_\alpha / \xi) x_3 \sinh{\xi(H-x_3)}
\nonumber \\
&+& (C_j \delta_{i3} + B_j \delta_{\alpha i} \ci \lambda_\alpha / \xi) (x_3-H) \cosh{\xi(H-x_3)}.
\end{eqnarray}
The coefficients, $\{B_j, C_j, B_{ij}, C_{ij}\}$,  are coupled through the transformed incompressibility condition. Inserting the transformed pressure and velocity fields (\ref{appeq:PressureSolution}--\ref{appeq:VelocitySolution}) into (\ref{appeq:TransformedStokes2}) yields the relations
\begin{eqnarray}
\label{appeq:IncompressibilityConditions1}
C_j &=& \xi H B_j + \xi B_{3j} + \ci \lambda_\beta C_{\beta j},
\\
B_j &=& - \xi H C_j + \xi C_{3j} + \ci \lambda_\beta B_{\beta j}.
\end{eqnarray}
These coefficients will now be found by applying the boundary conditions on $\hat{\mathcal{W}}_{ij}$.

\subsection{Transforming the boundary conditions}
\label{appsubsec:Transforming}

The boundary conditions (\ref{appeq:ImageIntegralBCS1}--\ref{appeq:ImageIntegralBCS3}) for the liquid film can be transformed by recognising that the integral form in which they appear is the (inverse) zeroth order Hankel transform.
Therefore, these transform to
\begin{eqnarray}
\label{appeq:BC1Transformed}
4 \pi \mu \hat{\mathcal{W}}_{ij} (x_3 = 0)
&=& - \ci (\delta_{i3} \delta_{j \alpha} + \delta_{i \alpha} \delta_{j3}) 
\frac{\partial}{\partial \lambda_\alpha} \left[ \frac{\sinh{\xi (H-h)}}{\sinh{\xi H}} \right]
\qquad \forall i,
\\
\label{appeq:BC2Transformed}
4 \pi \mu \hat{\mathcal{W}}_{ij} (x_3 = H)
&=& + \ci  \delta_{i3} \delta_{j \alpha} 
\frac{\partial}{\partial \lambda_\alpha} \left[ \frac{\sinh{\xi h}}{\sinh{\xi H}} \right]
\qquad \qquad \qquad \textmd{for } i=3,
\\
\label{appeq:BC3Transformed}
 \partial_3 4 \pi \mu \hat{\mathcal{W}}_{ij} (x_3 = H)
&=& \delta_{i\alpha} \delta_{j\beta} 
 \left( 2 \delta_{\alpha \beta} + \frac{\lambda_\alpha \lambda_\beta}{\xi} \frac{\partial}{\partial \xi} \right) \frac{\sinh{\xi h}}{\sinh{\xi H}}
\quad \textmd{for } i=1,2.
\end{eqnarray}
Here, it should be noted that the indices $\alpha, \beta \in \{1,2\}$, so for $j=3$ the right-hand sides of equations  (\ref{appeq:BC2Transformed}) and (\ref{appeq:BC3Transformed}) are equal to zero. 

\subsection{Solving the transformed auxiliary solution}
\label{appsubsec:Solving}

By equating these transformed boundary conditions to the ansatz (\ref{appeq:VelocitySolution}), the coefficients $\{B_{ij}, C_{ij}\}$ are determined to be
\begin{eqnarray}
\label{appeq:FourierSolved1}
B_{\alpha \beta} &=& \frac{1}{2\pi \xi^2} \left( 
-2 \delta_{\alpha \beta} \frac{\sinh{\xi h}}{\sinh{\xi H}} \xi +
2 \pi \ci (B_\beta - H \xi C_\beta) \lambda_\alpha 
- \frac{\partial}{\partial \xi} \left[ \frac{\sinh{\xi h}}{\sinh{\xi H}} \right] \lambda_\alpha \lambda_\beta
\right),
\nonumber \\
B_{\alpha 3} &=& \frac{\ci (B_3 - H \xi C_3) \lambda_\alpha}{\xi^2},
\nonumber \\
B_{3j} &=& H C_j \coth{\xi H}  + \ci \delta_{j\alpha} \frac{H-h}{2\pi \xi} \frac{\sinh{\xi h}}{\sinh{\xi H}} \lambda_\alpha,
\nonumber \\
C_{\alpha \beta} &=& \frac{\tanh{\xi H}}{2\pi \xi^2} \left( 
2 \delta_{\alpha \beta} \frac{\sinh{\xi h}}{\sinh{\xi H}} \xi 
+
2 \pi \ci (B_\beta (\xi H \coth{\xi H}- 1) + C_\beta \xi H ) \lambda_\alpha 
+ \frac{\partial}{\partial \xi} \left[ \frac{\sinh{\xi h}}{\sinh{\xi H}} \right] \lambda_\alpha \lambda_\beta
\right),
\nonumber \\
C_{\alpha 3} &=& \frac{\ci \lambda_\alpha \tanh{\xi H}}{4\pi \xi^2} \left( 
- \frac{2 \xi}{\sinh{\xi H}} \frac{\partial}{\partial \xi} \left[ \frac{\sinh{\xi (H-h)}}{\sinh{\xi H}} \right]
+ 4 \pi (B_3 (\xi H \coth{\xi H}- 1) + C_3 \xi H )
\right),
\nonumber \\
C_{3j} &=&
 \frac{\delta_{j\alpha}}{2\pi \xi} \frac{\partial}{\partial \xi} \left[ \frac{\sinh{\xi h}}{\sinh{\xi H}} \right] \ci \lambda_\alpha,
\end{eqnarray}
where the pressure coefficients are 
\begin{eqnarray}
\label{appeq:FourierSolved2}
B_j &=& - \ci \delta_{j\alpha} \lambda_\alpha \frac{1}{2 \pi \xi} \frac{\sinh{\xi h}}{\sinh{\xi H}}, 
\\
\label{appeq:FourierSolved3}
C_\alpha &=&  -\frac{\ci \lambda_\alpha}{2\xi H - \sinh{2\xi H}} \frac{1}{\pi \xi} \left( 
\xi h \sinh{\xi (H-h)} + \sinh{\xi h} \sinh{\xi H}
\right), 
\\
\label{appeq:FourierSolved4}
C_3 &=& - 
\frac{\sinh{\xi H}}{2\xi H - \sinh{2\xi H}}
\frac{\xi}{\pi}  \frac{\partial}{\partial \xi} \left[ \frac{\sinh{\xi (H-h)}}{\sinh{\xi H}} \right].
\end{eqnarray}
Inserting the coefficients (\ref{appeq:FourierSolved1}--\ref{appeq:FourierSolved4}) into equations (\ref{appeq:PressureSolution}--\ref{appeq:VelocitySolution}), we obtain complete expressions for the transformed velocity and pressure. 
These must be inverse transformed to obtain $\mathcal{W}_{ij}$ and $\mathcal{S}_j$, as described next. 

\subsection{Inverse transforming the auxiliary solution}
\label{appsubsec:Inverse}

The transformed auxiliary solution (\ref{appeq:PressureSolution}--\ref{appeq:VelocitySolution}) must now be inverse transformed.
We first consider $\mathcal{W}_{11}$, which has a term proportional to unity and another to $\lambda_1^2$. 
\begin{eqnarray}
\label{appeq:W11}
4\pi \mu \mathcal{W}_{11} &=& \int_{0}^\infty 
\frac{4 \sinh{\xi h} \sinh{\xi x_3}}{\xi \sinh{2 \xi H}}
J_0(\rho \xi) \xi d\xi 
+ \frac{\partial}{\partial r_1} \frac{r_1}{\rho} \int_{0}^\infty \hat{A}_{11}(\xi, x_3) J_1(\rho \xi) \xi^2 d\xi, \quad
\end{eqnarray}
where 
\begin{eqnarray}
\hat{A}_{11} &=&
-\frac{1}{\xi ^3 (\sinh (2 H \xi )-2 H \xi )}
\Big(
2 \sinh (\xi  (H-x_3))
\nonumber \\ && 
(\sinh (h \xi ) \sinh (H \xi ) (\coth (H \xi ) (H^2 \xi ^2 \textmd{csch}^2(H \xi )-1)
\nonumber \\ &&
+\xi  (x_3-2 H))+h \xi ^2 (H-x_3) \sinh (\xi  (h-H)))
\nonumber \\ &&
+\cosh (\xi  (H-x_3)) (2 \sinh (h \xi ) (\sinh (H \xi )
\nonumber \\ &&
+\xi  (2 H-x_3) \cosh (H \xi )
+H \xi  (\xi  (x_3-H) \textmd{csch}(H \xi )-2 \textmd{sech}(H \xi )))
\nonumber \\ &&
-2 h H \xi ^2 \tanh (H \xi ) \sinh (\xi  (h-H)))+2 h \xi  \cosh (h \xi ) 
\nonumber \\ &&
(2 H \xi  \textmd{csch}(2 H \xi )-1) \sinh (\xi  x_3)
\Big).
\end{eqnarray}
Similarly, with $\hat{A}_{11} = \hat{A}_{22}$, we have
\begin{eqnarray}
\label{appeq:W22}
4\pi \mu \mathcal{W}_{22} &=& \int_{0}^\infty 
\frac{4 \sinh{\xi h} \sinh{\xi x_3}}{\xi \sinh{2 \xi H}}
J_0(\rho \xi) \xi d\xi 
+ \frac{\partial}{\partial r_2} \frac{r_2}{\rho} \int_{0}^\infty \hat{A}_{22}(\xi, x_3) J_1(\rho \xi) \xi^2 d\xi. \quad
\end{eqnarray}
For $i=\alpha \neq j = \beta$, there is only one term proportional to $\lambda_\alpha \lambda_\beta$, so
\begin{eqnarray}
\label{appeq:W12}
4\pi \mu \mathcal{W}_{\alpha \beta} &=& \frac{\partial}{\partial r_\beta} \frac{r_\alpha}{\rho} \int_{0}^\infty \hat{A}_{\alpha \beta}(\xi, x_3) J_1(\rho \xi) \xi^2 d\xi,
\end{eqnarray}
where
\begin{eqnarray}
\hat{A}_{12} = \hat{A}_{21} &=&
- \frac{1}{\xi ^3 (\sinh 2 H \xi -2 H \xi )}
\Big(
2 \sinh (\xi  (H-x_3)) (\sinh (h \xi ) \sinh (H \xi ) 
\nonumber \\ &&
(\coth (H \xi ) (H^2 \xi ^2 \textmd{csch}^2(H \xi )-1)+\xi  (x_3-2 H))+h \xi ^2 (H-x_3) \sinh (\xi  (h-H)))
\nonumber \\ &&
+\cosh (\xi  (H-x_3)) (2 \sinh (h \xi ) (\sinh (H \xi )+\xi  (2 H-x_3) \cosh (H \xi )
\nonumber \\ &&
+H \xi  (\xi  (x_3-H) \textmd{csch}(H \xi )-2 \textmd{sech}(H \xi )))-2 h H \xi ^2 \tanh (H \xi ) \sinh (\xi  (h-H)))
\nonumber \\ &&
+2 h \xi  \cosh (h \xi ) (2 H \xi  \textmd{csch}(2 H \xi )-1) \sinh (\xi  x_3)
\Big).
\end{eqnarray}
For $i=3, j = \alpha$ there is only one term proportional to $\lambda_\alpha$, so
\begin{eqnarray}
\label{appeq:W31}
4\pi \mu \mathcal{W}_{3 \alpha} &=& \frac{r_\alpha}{\rho} \int_{0}^\infty \hat{A}_{3 \alpha}(\xi, x_3) J_1(\rho \xi) \xi^2 d\xi,
\end{eqnarray}
where
\begin{eqnarray}
\hat{A}_{31} = \hat{A}_{32} &=&
\frac{1}{\xi } h \frac{\cosh (h \xi )}{\sinh(\xi H)} \cosh (\xi  (H-x_3))
\nonumber \\ 
&+& \frac{1}{\xi } \frac{1}{2 H \xi -\sinh (2 H \xi )} \frac{1}{\sinh(\xi H)}
\Big(
\cosh (\xi  (H-x_3)) (\sinh (h \xi ) (-2 H^2 \xi  \coth (H \xi )
\nonumber \\ &&
+(2 H-x_3) \cosh (2 H \xi )+x_3)+2 h \xi  (x_3-H) \sinh (H \xi ) \sinh (\xi  (h-H)))
\nonumber \\ &&
+\sinh (\xi  (H-x_3)) (2 h H \xi  \cosh (H \xi ) \sinh (\xi  (h-H))
\nonumber \\ &&
+\sinh (h \xi ) ((h-2 H+x_3) \sinh (2 H \xi )-2 H \xi  (h-H+x_3)))
\Big).
\end{eqnarray}
For $i=\alpha, j = 3$, there is also only one term proportional to $\lambda_\alpha$, so 
\begin{eqnarray}
\label{appeq:W13}
4\pi \mu \mathcal{W}_{\alpha 3} &=& \frac{r_\alpha}{\rho} \int_{0}^\infty \hat{A}_{\alpha 3} (\xi, x_3) J_1(\rho \xi) \xi^2 d\xi,
\end{eqnarray}
where
\begin{eqnarray}
\hat{A}_{13} = \hat{A}_{23} &=&
\frac{1}{\xi } \frac{1}{2 H \xi -\sinh (2 H \xi )} \frac{1}{\sinh(\xi H)}
\Big(
((h-2 H) \sinh (h \xi )-h \sinh (\xi  (h-2 H)))
\nonumber \\ &&
(\xi  (x_3-H) \sinh (\xi  (H-x_3))+(H \xi  \coth (H \xi )-1) \cosh (\xi  (H-x_3)))\Big).
\end{eqnarray}
For $i= j = 3$,
\begin{eqnarray}
\label{appeq:W33}
4\pi \mu \mathcal{W}_{33} &=& \int_{0}^\infty \hat{A}_{33} (\xi, x_3) J_0(\rho \xi) \xi d\xi,
\end{eqnarray}
where
\begin{eqnarray}
\hat{A}_{33} &=&
\frac{\xi}{2 H \xi -\sinh (2 H \xi )} \frac{1}{\sinh(\xi H)}
\Big(
((h-2 H) \sinh (h \xi )-h \sinh (\xi  (h-2 H))) 
\nonumber \\ &&
((x_3-H) \cosh (\xi  (H-x_3))+H \coth (H \xi ) \sinh (\xi  (H-x_3)))
\Big).
\end{eqnarray}
Finally, the auxiliary pressure field is given by
\begin{eqnarray}
\label{appeq:AuxiliaryPressure}
\mathcal{S}_j &=& 
 \int_{0}^\infty \left[
\delta_{\alpha j} 
\frac{r_\alpha}{\rho} \hat{A}_\alpha (\xi, x_3) J_1(\rho \xi) \xi^2 
+ \delta_{3 j} 
\hat{A}_3 (\xi, x_3) J_0(\rho \xi) \xi 
\right] d\xi,
\end{eqnarray}
where
\begin{eqnarray}
\hat{A}_1 = \hat{A}_2 &=&
\frac{1}{\pi \xi}
\frac{ \cosh \xi  (H-x_3)}{2 H \xi -\sinh (2 H \xi )}
\Big(
h \xi  \sinh (\xi  (h-H))-\sinh (h \xi ) \sinh (H \xi )
\Big)
\nonumber \\
&-& \frac{1}{2 \pi \xi} \frac{\sinh{\xi h}}{\sinh{\xi H}} \sinh{\xi (H-x_3)},
\\
\hat{A}_3 &=&
\frac{\xi}{2\pi}
\frac{1}{2 H \xi -\sinh (2 H \xi )} \frac{\cosh \xi  (H-x_3)}{\sinh\xi H}
\nonumber \\ && \cdot
\Big(
-2 H \sinh (h \xi )-h \sinh (\xi  (h-2 H))+h \sinh (h \xi )
\Big).
\end{eqnarray}
The final solution for the flow $\vec{u}^\textmds{F}$ and pressure $P^\textmds{F}$ generated by a Stokeslet of force strength $\vec{f}$ in a liquid film is then obtained by adding the expressions for the image series (\ref{appeq:ImageIntegral}--\ref{appeq:ImageIntegralPressure}) and the auxiliary solution (\ref{appeq:W11}--\ref{appeq:AuxiliaryPressure}), so that 
\begin{eqnarray}
\label{appeq:FTFinalSolution}
u_i^\textmds{F} = \mathcal{F}_{ij} f_j \quad \textmd{where} \quad \mathcal{F}_{ij} = \mathcal{V}_{ij} + \mathcal{W}_{ij},
\end{eqnarray}
and
\begin{eqnarray}
\label{appeq:FTFinalSolutionPressure}
P^\textmds{F} = \mathcal{P}_{j} f_j \quad \textmd{where} \quad \mathcal{P}_j = \mathcal{Q}_j + \mathcal{S}_j.
\end{eqnarray}


\section{Velocity and pressure in the thin-film limit}
\label{appsec:FarField}

In this section, we consider the solution outlined in Appendix \sect{appsec:StokesletFlowFilm} in the thin-film limit where the film height is much smaller than the lateral distances between the flow source and the point where the flow is evaluated ($H \ll \rho$).
This limit can also be seen as the far-field limit when all distances are large compared to the film height.
To achieve this result, we follow \citet{liron1976Stokes} further, and transform the integral expressions into an alternative form. 
Once we have obtained the thin-film expression for the Stokeslet, we continue to derive the thin-film flows and pressures generated by a force-free and torque-free micro-swimmer through the use of a multipole expansion (\sect{sec:multipoleExpansion}). 
This could be particularly relevant in the study of hydrodynamically interacting microbes with a size comparable to the film height, when neighbouring organisms are a distance of more than one film height apart in the $\rho$ direction.

\subsection{Stokeslet in a thin liquid film}
\label{appsubsec:FarFieldStokesletFilm}

To transform the integral expressions in equations (\ref{appeq:ImageIntegral}--\ref{appeq:ImageIntegralPressure}) and (\ref{appeq:W11}--\ref{appeq:AuxiliaryPressure}), we use the Hankel transformation \citep[][p 296]{liron1976Stokes}, given by
\begin{eqnarray}
\label{appeq:HankelTransformation}
\int_0^{\infty} J_\nu (b \xi) \xi^{\nu+1} F(\xi) d\xi = \ci \pi \textmd{(sum of residues}
\nonumber \\ 
\textmd{in the upper half plane of $F(z)z^{\nu+1}H_{\nu}^{(1)}(b z)$}
\nonumber \\ 
\textmd{including one half of the residue at $z=0$)},
\end{eqnarray}
where $b$ is real, $F(z)$ is an even function of $z$ that decays exponentially to zero on the real axis as $\textmd{Re } z \to \pm \infty$, and the integral is taken over the Hankel contour $C$ in the complex plane \citep[][Figure 3]{liron1976Stokes}. 
Using this Hankel transformation, the integrals can be written as infinite series
\begin{eqnarray}
\label{appeq:HankelTransformation1}
\int_0^{\infty} J_0 (\rho \xi) \frac{\sinh{\xi h}}{\sinh \xi H} \sinh{\xi(H-x_3)} d\xi = \frac{2}{H} \sum_{n=1}^\infty \sin \frac{n \pi h}{H} \sin \frac{n \pi x_3}{H} K_0 \left ( \frac{n \pi \rho}{H} \right ),
\\
\label{appeq:HankelTransformation2}
\int_0^{\infty} \xi  J_1 (\rho \xi) \frac{\sinh{\xi h}}{\sinh \xi H} \sinh{\xi(H-x_3)} d\xi = \frac{2}{H} \sum_{n=1}^\infty \frac{n \pi}{H} \sin \frac{n \pi h}{H} \sin \frac{n \pi x_3}{H} K_1 \left ( \frac{n \pi \rho}{H} \right ).
\end{eqnarray}
By identifying the singularities in the functions $\hat{A}_{ij}$, the residues can be found and hence the integrals in equations  (\ref{appeq:W11}--\ref{appeq:W33}) can be transformed.

For a Stokeslet bounded between a no-slip wall and a parallel no-shear interface, this leads to a far-field velocity field $\mathcal{F}_{ij}$ that decays exponentially with $\rho$ if either $i$ or $j$ or both are equal to three. 
Therefore, the only components of the flow that do not decay exponentially are those oriented parallel to the surfaces, if the Stokeslet is also oriented parallel to the surfaces. These are flows with a half-parabolic profile with a maximum at the top interface 
\begin{eqnarray}
\label{appeq:StokesletFilmFF}
\mathcal{F}_{ij} \sim - \frac{3H}{\pi \mu} \frac{x_3}{H} \left( 1- c \frac{x_3}{H} \right)  \frac{h}{H} \left( 1- c \frac{h}{H} \right) \frac{1}{\rho^2} \left [ \frac{\delta_{\alpha \beta}}{2} - \frac{r_\alpha r_\beta}{\rho^2}  \right ] \delta_{i \alpha} \delta_{j \beta} + \mathcal{O} \left( e^{- \rho/H} \right),
\end{eqnarray}
where $c=1/2$. 
The structure of this flow in the $x_1-x_2$ plane is shown in \fig{appfig:ThinFilmFlows}a.
The corresponding pressure due to a Stokeslet in a thin film is given by
\begin{eqnarray}
\label{appeq:StokesletPressureFilmFF}
\mathcal{P}_j \sim \frac{3}{2 \pi H} \frac{h}{H} \left( 1- c  \frac{h}{H} \right) \frac{r_\alpha}{\rho^2} 
\delta_{j \alpha} + \mathcal{O} \left( e^{- \rho/H} \right).
\end{eqnarray}

The same far-field analysis can be performed for a Stokeslet bounded between two parallel plates \citep{liron1976Stokes}.
The $r_1$ and $r_2$ dependence of these solutions are identical, and the solution is obtained by simply setting $c=1$. 
Both thin-film/far-field solutions can be classified as having the structure of a two-dimensional source doublet. 
That is, if a 2D source generates a flow $s_i = r_i/\rho^2$, then a 2D source doublet oriented in the $j$ direction generates a flow $d_{ij} = - \partial_j s_i = - \delta_{ij} / \rho^2 + 2 r_i r_j / \rho^4$. 

\subsection{Micro-swimmer in a thin liquid film}
\label{appsubsec:FarFieldStokesletFilm}

\begin{figure}
	\includegraphics[width = \linewidth]{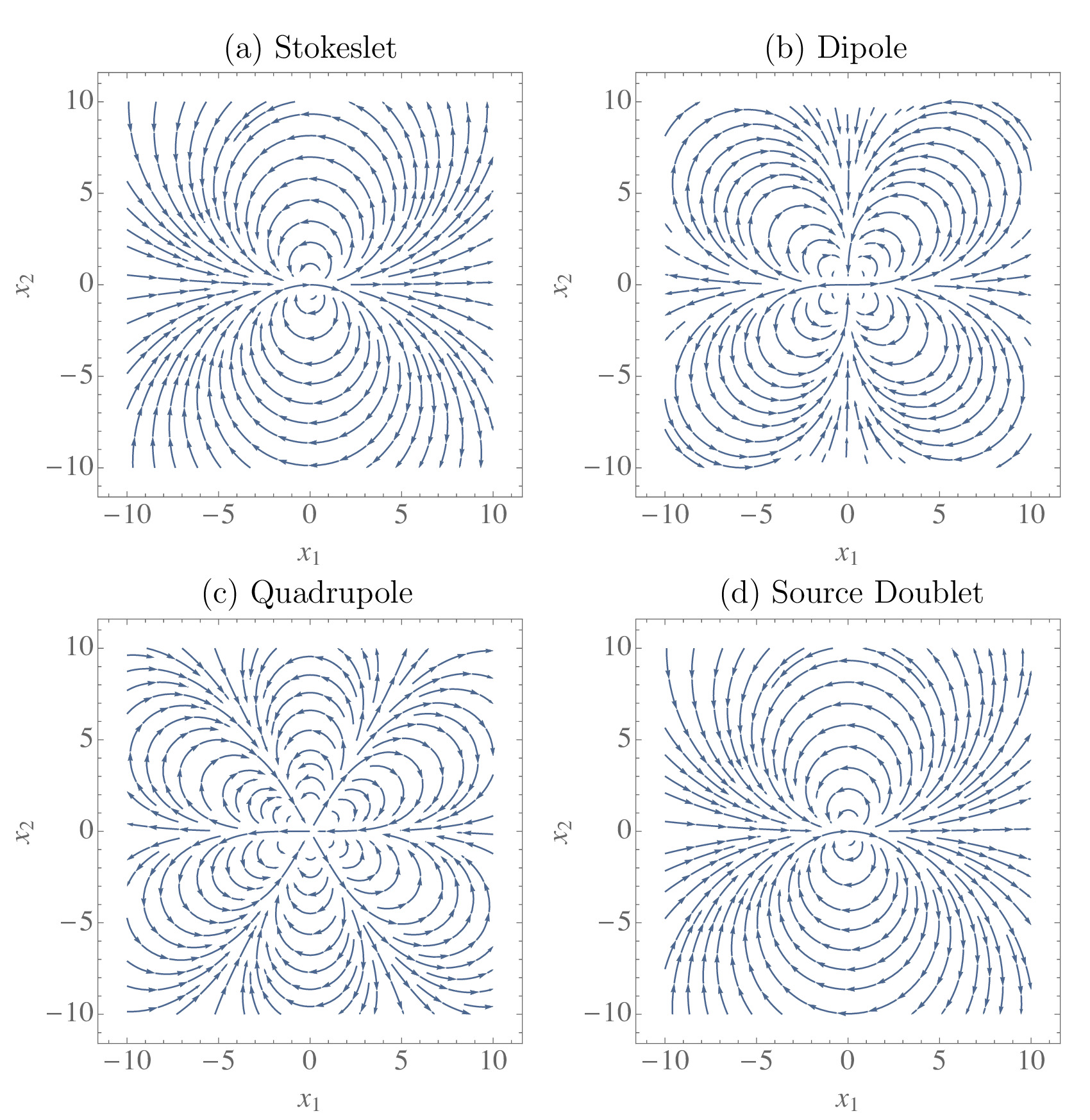}
	\caption{Flow fields generated by a micro-swimmer in the thin-film limit, equivalent to the limit far from the swimmer in the lateral direction, so that $H \ll \rho$. The swimmer is oriented parallel to the film surfaces, in the positive $x_1$ direction. Shown are streamlines in the $x_3 = H$ plane for the (a) Stokeslet, (b) Stokes dipole, (c) quadrupole and (d) source doublet.}
	\label{appfig:ThinFilmFlows}
\end{figure}

We now extend this result to deduce the flows generated by a force-free and torque-free micro-swimmer in a thin liquid film. We do this by expressing the swimmer-generated flows by a multipole expansion, as introduced in \sect{sec:multipoleExpansion}, written as
\begin{eqnarray}
 \label{appeq:swimmerVelocityDef}
 \vec{u} \left(\vec{x}, \vec{y}, \vec{p} \right) &= \vec{u}^\textmds{D} + \vec{u}^\textmds{Q} + \vec{u}^\textmds{SD} + \vec{u}^\textmds{RD} + \ldots ,
\end{eqnarray}
where the swimmer is located at the position $\vec{y}$ and oriented in the direction $\vec{p}$, and the multipole contributions are the Stokes dipole $\vec{u}^\textmds{D}$, the quadrupole $\vec{u}^\textmds{Q}$, the 3D source doublet $\vec{u}^\textmds{SD}$, and the rotlet doublet $\vec{u}^\textmds{RD}$.
Each contribution to the multipole expansion can be written in terms of derivatives of the Green's function in the thin film (\eq{appeq:StokesletFilmFF}), 
as given by \textbf{Eqs.~\ref{eq:flowStokesDipole}--\ref{eq:flowRotletDipole}}. 

If we look at flows parallel to the film surfaces ($i = 1,2$) and the swimmer is also oriented parallel to the surfaces ($\vec{p} = \hat{\vec{e}}_j$ with $j = 1,2$), then the leading dipolar contribution to the flow field far from the swimmer is given by 
\begin{eqnarray}
\label{appeq:FarFieldFlowStokesDipole}
u_i^{\textmds{D}}(\vec{x}, \vec{y}, \vec{\hat{e}}_j) &=&
- 24 \kappa H \frac{x_3}{H} \left( 1- c \frac{x_3}{H} \right)  \frac{h}{H} \left( 1- c \frac{h}{H} \right)
\left [ \frac{2 \delta_{i j} r_j}{\rho^4} +  \frac{r_i}{\rho^4} - \frac{4 r_i r_j^2}{\rho^6}  \right ],
\end{eqnarray}
with $c=1/2$ for the film. 
The quadrupole term is 
\begin{eqnarray}
\label{appeq:FarFieldFlowStokesQuadrupole}
u_i^{\textmds{Q}}(\vec{x}, \vec{y}, \vec{\hat{e}}_j) &=&
- 36 \nu H \frac{x_3}{H} \left( 1- c \frac{x_3}{H} \right)  \frac{h}{H} \left( 1- c \frac{h}{H} \right)
\left [ \frac{\delta_{i j}}{\rho^4} -  \frac{4  \delta_{i j} r_j^2}{\rho^6} -  \frac{4  r_i r_j}{\rho^6} + \frac{8 r_i r_j^3}{\rho^8}  \right ], \qquad \quad
\end{eqnarray}
and the 3D source doublet is 
\begin{eqnarray}
\label{appeq:FarFieldFlowSourceDoublet}
u_i^{\textmds{SD}}(\vec{x}, \vec{y}, \vec{\hat{e}}_j) &=&
- 24c \sigma \frac{x_3}{H^2} \left( 1- c  \frac{x_3}{H} \right)
\left [ \frac{\delta_{i j}}{2 \rho^2} - \frac{r_i r_j}{\rho^4}  \right ].
\end{eqnarray}

For flows ($i=3$) or swimmers ($j=3$) oriented perpendicular to the film surfaces, the dipolar, quadrupolar and source dipole flow fields decay exponentially in the far field. 

\fig{appfig:ThinFilmFlows} shows the flow fields generated by a Stokeslet (a) or by the dipole and higher-order multipoles of a force- and torque-free microswimmer (b--d).
The Stokeslet flow consists of a recirculating flow pattern of two loops, whereas the dipole has four loops, the quadrupole has six loops, and the source doublet maintains its bulk-flow structure with two loops.
These recirculating flow patterns are an effect of the minimisation of energy dissipation of Stokesian flow, because the energy loss by vortices (shear gradients) becomes less and less significant compared to viscous dissipation by the boundaries with increasing confinement.

\subsection{Micro-swimmer between parallel plates}
\label{appsubsec:FarFieldStokesletFilm}

For completeness, we also give the flow fields generated by a micro-swimmer in a channel composed of two closely spaced parallel flat plates. The structure of these flows is identical to those in a thin liquid film (\ref{appeq:FarFieldFlowStokesDipole}--\ref{appeq:FarFieldFlowSourceDoublet}), but using the prefactor $c=1$.

\section{Comparison between the recursive series and the exact solution}
\label{appsec:Comparison}

\begin{figure}
	\begin{center}
    	\includegraphics[width=\linewidth]{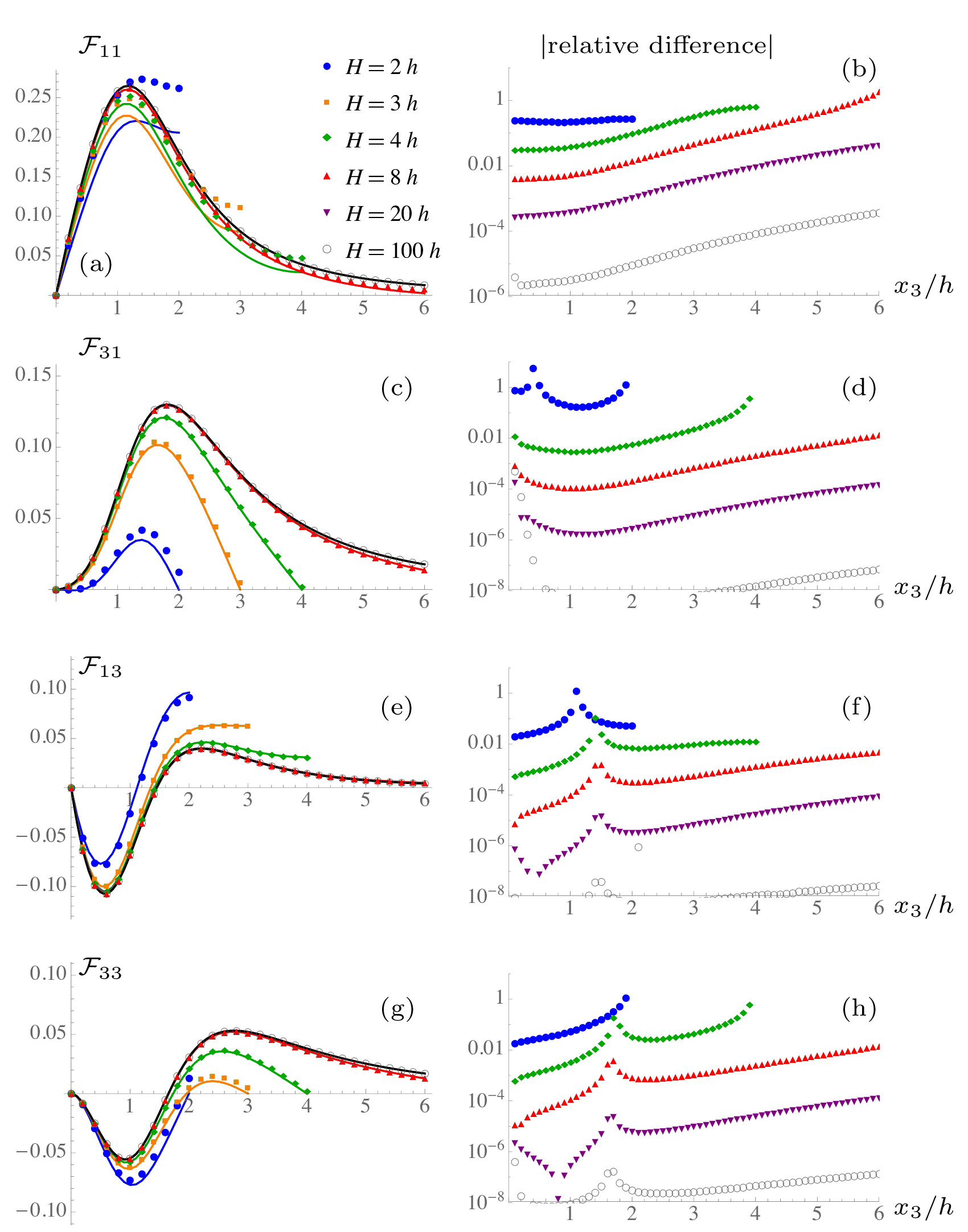}
	\end{center}
  	\caption{
Comparison of the velocity profiles of the Stokeslet flow $u_i^\textmd{S}(\vec{x}, \vec{y}, f \vec{\hat{e}}_j) = \mathcal{F}_{ij} f$ for various values of the film height $H$, as a function of the dimensionless coordinate $x_3/h$, with a constant Stokeslet position $y_3 = h$ and constant $r_1 = r_2 = h$.
The Stokeslet is oriented parallel (a--d; $\vec{f} = f \hat{\vec{e}}_1$) or perpendicular (e--h; $\vec{f} = f \hat{\vec{e}}_3$) to the film surfaces, and has force strength $f=4 \pi \mu$.
The left panels show flows obtained with the exact solution (solid lines), and with the recursive series method (points), where $n=9$ images have been used.
The right panels show the relative difference between the values obtained from the two methods.
The solid black line represents the case of $H \to \infty$, where only one no-slip wall is present, plotted using the Blake solution (\eq{eq:blakeTensor1}). 
}
  	\label{appfig:LironPlots1}
\end{figure}

In \fig{appfig:LironPlots1} we show the velocity profiles generated by a Stokeslet for various film heights $H$. 
We compare the solutions obtained with the recursive series method (\sect{sec:filmStokeslet}) and with the exact solution (Appendix \sect{appsec:StokesletFlowFilm}).
The left panels show the absolute flow values for both solutions (points and solid lines, respectively), and the right panels show their relative difference.
Note that the same parameters have been chosen as in \citet{liron1976Stokes} so that a direct comparison can also be made with \textbf{Figs. 4--6} therein.

The flow structure is displayed in the left panels of \fig{appfig:LironPlots1}. 
Flows parallel to the film surfaces due to a Stokeslet also oriented parallel, (\fig{appfig:LironPlots1}a), are half-parabolic in the thin-film limit, as discussed in Appendix \sect{appsec:FarField}.
For film heights comparable to the swimmer height (blue dotted line; $H = 2 h$) the no-shear condition is still satisfied at the top interface, and the flow is still half-parabolic near the bottom wall. The overall profile however is non-trivial.
In the thick-film limit (black line; $H \to \infty$) the single-wall result is recovered.
Flows oriented perpendicular to the surfaces (\fig{appfig:LironPlots1}c,g) are not half-parabolic, but vanish at both $x_3 = 0,H$.
If the Stokeslet is oriented parallel to the surfaces (\fig{appfig:LironPlots1}a,c), the values do not cross zero, whereas for Stokeslets oriented perpendicular (\fig{appfig:LironPlots1}e,g) they do to satisfy incompressibility.
In agreement with the conclusion by \citet{liron1976Stokes}, if $H< 8h$, we find that the effect of the second surface cannot be neglected.

Lastly, we discuss the relative difference between the two solutions, defined as $|(\mathcal{F}_{ij}^\textmds{series}-\mathcal{F}_{ij}^\textmds{exact})/\mathcal{F}_{ij}^\textmds{exact}|$, and thus the accuracy of the recursive series solution (\fig{appfig:LironPlots1}, right panels).
The agreement is not acceptable in the thin-film limit, $h \sim r_1 < H$, as more images are required because they become approximately equidistant to the point of interest.
However, the agreement is good for large films heights, $h \sim r_1 \ll H$, where the series can be safely truncated.
For example, if $H = 100h$ the error is $<10^{-4}$ for all $x_3$ values, with the largest error at the top interface and the smallest error at the bottom surface as the series solution satisfies the no-slip boundary conditions exactly with $n=9$ images.


\bibliographystyle{jfm}
\bibliography{lit}
\end{document}